\documentclass[entropy,article,accept,oneauthor,pdftex,12pt,a4paper]{mdpi}
\lastpage{1832}
\doinum{10.3390/e12071799}
\pubvolume{12}
\pubyear{2010}
\history{Received: 5 June 2010; in revised form: 5 July 2010 / Accepted: 9 July 2010 / \linebreak Published: 20 July 2010}

\usepackage{amssymb,amsthm}
\usepackage{caption}
\usepackage{hyphenat}
\usepackage[sort&compress]{natbib}
\usepackage{hypernat}
\newcommand{\T}{{\rm Tr}}

\newcommand{\cB}{{\cal B}}
\newcommand{\cH}{{\cal H}}

\newcommand{\cL}{{\cal L}}

\newcommand{\0}{{\mathbf 0}}
\newcommand{\1}{{\mathbf 1}}

\def\bbbr{{\rm I\!R}}
\def\bbbc{{\mathchoice {\setbox0=\hbox{$\displaystyle
\rm C$}\hbox{\hbox
to0pt{\kern0.4\wd0\vrule height0.9\ht0\hss}\box0}}
{\setbox0=\hbox{$\textstyle\rm C$}\hbox{\hbox
to0pt{\kern0.4\wd0\vrule height0.9\ht0\hss}\box0}}
{\setbox0=\hbox{$\scriptstyle\rm C$}\hbox{\hbox
to0pt{\kern0.4\wd0\vrule height0.9\ht0\hss}\box0}}
{\setbox0=\hbox{$\scriptscriptstyle\rm C$}\hbox{\hbox
to0pt{\kern0.4\wd0\vrule height0.9\ht0\hss}\box0}}}}

\Title{Roofs and Convexity}

\Author{Armin Uhlmann}

\address{%
Institute for Theoretical Physics, University of Leipzig,
PB 100 920, D-04009 Leipzig, Germany; E-Mail: armin.uhlmann@t-online.de
}

\corres{armin.uhlmann@itp.uni-leipzig.de}

\abstract{ Convex roof extensions are widely used to create
entanglement measures in quantum information theory. The aim of the
article is to present some tools which could be helpful for their
treatment. Sections 2 and 3 introduce into the subject. It follows
descriptions of the Wootters' method, of the ``subtraction
procedure'', and examples on how to use symmetries. }


\begin{document}

\section{Introduction}
One often is in the position to know a quantity for pure states of
a quantum system, say $g$, without a definite meaning in classical
physics. Then one looks for a method to extend $g$ to all mixed
states, \linebreak see \cite{NC00,Vi02,BZ06,HHHH07}. Let $G$
denote such a possible extension.

A reasonable approach is certainly to extend $g$ ``as linearly as
possible'' or, more correctly, ``as affine as possible''. The
mathematical model for such a demand is a ``roof'' as presented in
Section 2.

As it is difficult, to say the least, to imagine higher dimensional
geometry, let us look at an elementary example, the real qubits.
They fill a disk bounded by a circle. The circle represents the pure
states. Considering $g(\pi)$, with $\pi$ pure, as the height of a
wall at the point $\pi$. An extension $G$ of $g$ to the disk
provides a covering of the ground floor. Let $G(\omega)$ be a point
of that covering. To be a roof in the sense of Section 2, we like to
have: There is either a straight line or a plane coincident with the
roof point $\omega, G(\omega)$ and resting on the wall at two or at
three points $\pi_j, g(pi_j)$. If the line or plane is parallel to
the ground floor, it is called {\em flat}. An uncomfortable feature
is the rather large arbitrariness: There are plenty of roof
extensions allowing, however, as a bonus, much room for playing.
Under rather weak assumptions, there is a minimal as well as a
maximal roof extension (Proposition 2.1).

Physically stronger motivated seem extensions $G$ of $g$ which are
either convex or concave. Heuristically, the convex ones try to
suppress the ``classical noise'' or else the ``classical
information'' by the tendency to attach lower values to the states
``far'' from the pure ones. The concave extension stress the
``classical part'', possessing lower values in the vicinity of the
pure states.

It is a well known fact that there are maximal convex and minimal
concave extensions. If it exists, the minimal roof extension is
equal to the maximal convex extension and this facilitate \linebreak
calculations sometimes.

The maximal convex extension $g^{\cup}$ of $g$ can be gained
by
\begin{equation} \label{0.1}
g^{\cup}(\omega) = \inf \sum p_j g(\pi_j)
\end{equation}
where the inf is running through all convex decompositions
\begin{equation} \label{0.2}
\omega = \sum p_j \pi_j, \quad p_j \geq 0. \quad \sum p_j =1
\end{equation}
with pure states $\pi_j$, \cite{Roc70}. In quantum information
theory this is nowadays a common procedure to define entanglement
measures. Its first appearance is in the important paper
\cite{BDSW96} by C.~H. Bennett, D.~Di{V}incenzo, J.~Smolin, and
W.~Wootters, and has been called {\em entanglement of formation} and
is denoted by $E(\omega)$. These authors considered states $\omega$
of a bipartite quantum system. The role of $g$ plays $S(\T_b \pi)$,
the entropy $S$ of the partial trace $\T_b$, as a function of a pure
bipartite states $\pi$.

In the present paper a trace preserving positive map $T$ from the
states of a quantum system into another one is called a {\em
stochastic map.} A {\em channel} is a completely positive stochastic
map. If in the definition of entanglement of formation, $E$, a
stochastic channel $T$ is used in place of the partial trace $\T_b$,
we call the resulting quantity {\em the entanglement of $T$} and
denote it by $E_T$. For a channel, $E_T$ is equivalent (in many
ways) to the restriction of $E$ onto a face of a bipartite state
space of sufficiently high dimension, \cite{HLW07}. In this way, $T$
is seen as a sub-channel of the partial trace: The tensor product is
partitioned into subspaces on which the sub-channels are defined. A
1-qubit channel, for example, can be represented by a $2 \times m$
bipartite quantum systems as the restriction of the partial trace
over the larger dimensional part onto the density operators
supported by a suitable 2-dimensional subspace.

Clearly, $E_T$ is of interest in its own as part of the
$\chi^*$-function $\chi^* = S_T - E_T$ with $(S_T)(\omega) =
S(T(\omega))$ the maximum of which is the Holevo capacity
\cite{Ho73} of $T$.

Since the importance of $E$ has been realized in \cite{BDSW96},
several other measures of similar structure have been introduced and
discussed, replacing the von Neumann entropy $S$ by another function
on the state space of the output system. The perhaps mostly
discussed examples are the ``tangle'' $\tau_T$ and the
``concurrence'' $C_T$. The connection between them is $C_T(\pi)^2 =
\tau_T(\pi)$ for all pure states $\pi$. Sometimes one also needs a
minimal concave extension (``entanglement of assistance''). All
these quantities have been defined by global variational problems of
type (\ref{0.1}), (\ref{0.2}).

A further, even earlier source for the said procedure roots in the
problem of defining a ``quantum dynamic entropy'', generalizing the
Kolmogorov--Sinai one. In the approach \cite{CNT87} of A.~Connes,
H.~Narnhofer, and W.~Thirring several similar global variational
problems wait to be solved. One of them is the search for the convex
roof defined by $g(\omega) = S(D_n(\omega))$, $S$ again von
Neumann's entropy and $D_n$ the diagonal map, setting all
off-diagonal entries of a matrix to zero. The problem initiated the
paper \cite{BNU96} of F.~Benatti, H.~Narnhofer, and A.~Uhlmann  and
further ones.

It is true that complete solutions for these extensions are only
known in the lowest non-trivial dimensions. In the present paper
general tools are presented to facilitate the treatment of convex or
concave roofs. The general aspects are mainly in the Sections 2 and
3. It includes the foliation of the input state space into ``convex
leaves'' onto which the  roof becomes affine and, in particular nice
situations, even constant, \cite{BNU96,Uh98a}.

Section 4 is devoted to the Wootters' way
\cite{HilWoo97,Woo97,Uh00c} of presenting the entanglement of
formation explicitly, see also \cite{Woo01}.

Section 5 shows a more recent way to compute the concurrence of a
1-qubit stochastic map by a substraction procedure. For the
concurrence of $2 \times m$ quantum systems it allows to compute the
concurrence of any rank two quantum state. An elegant way to do so
was opened by Hildebrand, \cite{hildebrand07,hildebrand08}, using
the so-called ``S-lemma''of Yakobovich, see \cite{PT07}. Another one
has been chosen by Hellmund and Uhlmann, \cite{HU08a}, who use the
description of general 1-qubit maps given by Gorini and Sudershan,
\cite{GS76}.

For the tangles the pioneering work goes back to Coffman, Kundu, and
Wootters, \cite{CKW00}, who already remarked that optimal
decompositions of length two should be sufficient in the 1-qubit
case. That the ``substraction procedure'' works can be read off from
a paper of Osborne and Verstraete, \cite{OV06}. Here we describe
analytical results for axial symmetric channels. There are also
results for the 3-tangle \linebreak roof problem, \cite{OSU08}.

Most results, if not numerically, are found by the help of
symmetries. Besides the already quoted ones, an essential step has
been done by K.~G.~Vollbrecht and R.~F.~Werner, \cite{VW96}, and
B.~M~.Terhal and K.~G.~Vollbrecht \cite{TV00}. Meanwhile it became a
very large domain of research, exceeding the frame of the present
paper. Hence, in Section 6, only some aspects, connected with
maximal symmetric states, \linebreak are touched.

Some notations: We use $\cH$ for Hilbert spaces, $\cB(\cH)$ for its
algebra of operators, $\Omega(\cH)$ for the set of density operators
supported by $\cH$. We also say ``state'' for ``density operator''
and ``state space'' for $\Omega(\cH)$. As a convex set,
$\Omega(\cH)$ is embedded in Herm$(\cH)$, the real linear space of
Hermitian operators. The symbol $\Omega$ is also used for a general
compact convex sets in a real linear space. Following \cite{BZ06},
the extremal points of a convex set will be called ``pure'' ones
mostly. They are usually symbolized by the letter $\pi$. We follow
\cite{OhyaPetz} in using $\eta(x) = -x \log x$ and $S(\omega) = \T
\, \eta(\omega)$, the von Neumann entropy.

\section{Roofs, Roof Extensions}
We are now going to give an exact meaning to the word ``roof''. For
this purpose we assume $G$ to be a real valued function on a compact
convex set $\Omega$, contained in a finite dimensional real linear
space.
\\[12pt]
\noindent {\bf Definition 2.1a: Roof points}
\vspace{12pt}

$\omega \in \Omega$ is called a {\em roof point of $G$,} if there is
at least one extremal convex decomposition
\begin{equation} \label{n2.2}
\omega = \sum p_j \pi_j, \quad \pi_j \in \Omega^{\rm pure}
\end{equation}
such that
\begin{equation} \label{n2.2f}
G(\omega) = \sum p_j G(\pi_j) \;
\end{equation}
If this takes place, we call the decomposition (\ref{n2.2}) {\em
optimal} with respect to $G$ or, equivalently, $G${\em -optimal.}
The number of terms in (\ref{n2.2}) with $p_j \neq 0$ is the {\em
length} of the decomposition.
\\[12pt]
\noindent {\bf Definition 2.1b: Flat roof points}
\vspace{12pt}

A roof point $\omega$ of $G$ is called {\em flat,} if there
is a $G$-optimal decomposition (\ref{n2.2}) fulfilling
\begin{equation} \label{n2.}
G(\omega) = G(\pi_1) = G(\pi_2) = \dots
\end{equation}
\emph{i.e.}~all the values $G(\pi_j)$ are equal one to another
\cite{fnote8}.

Let $f(x)$ be a real function defined on the range of $G$. The main
merit of a flat roof point $\omega$ of $G$ is the simple fact, that
it remains a flat roof point for $f(G)$: $f(G(\pi_j)) =
f(G(\omega))$ for all $j$ and (\ref{n2.2f}) remains true for $f(G)$.
In other words, the flat points of $\rho \to G(\rho)$ are also flat
roof points of $\rho \to f(G(\rho))$.
\\[12pt]
\noindent {\bf Definition 2.2: Roofs, flat roofs}
\vspace{12pt}

A real function $G$ on $\Omega$ is a {\em roof } if every
$\omega \in \Omega$ is a roof point of $G$.

$G$ is called a {\em flat roof } if all roof points are flat ones.

In important applications the point of view is a bit different: A
function $\pi \to g(\pi)$ is given on the set $\Omega^{\rm pure}$ of
pure states, or, more generally, on the set of extremal states of an
arbitrary compact convex set. Then one asks for meaningful
extensions $G$ which coincides with $g$ on $\Omega^{\rm pure}$. In
such a generality the problem is too arbitrarily posed and one asks
for restrictions to such an extension. Remarkable ones are the roof
extensions. They interpolate the values of $g$ ``as linearly (or as
affine) as possible''.
\\[12pt]
\noindent {\bf Definition 2.3: Roof extensions}
\vspace{12pt}

Let $\pi \to g(\pi)$ be a real function on $\Omega^{\rm pure}$.
A roof $G$ is called a {\em roof extension of $g$,} if
$G(\pi) = g(\pi)$ for pure states.

Now observe the following simple fact: The maximum (the minimum)
\begin{displaymath}
\max \{G_1, \dots, G_n \} \, \hbox{ respectively } \,
\min \{G_1, \dots, G_n \}
\end{displaymath}
of finitely many roof extensions of $g$ is a roof again.

Indeed, assume for roof extensions $G_1, \dots, G_n$ of $g$
and $\omega \in \Omega$ the value $G_1(\omega)$ is not less
than the other values $G_j(\omega)$. Then one selects a
$G_1$-optimal decomposition for
$\max \{G_1, \dots, G_n\}$.

The reasoning above fails for infinitely many roof extensions. The
proof of the following is postponed to that of proposition 3.5.
\\[12pt]
\noindent {\bf Proposition 2.1}
\vspace{12pt}

Let $g$ be a real continuous function on $\Omega^{\rm pure}(\cH)$
and $G$ a roof extension of $g$. For all $\omega \in \Omega(\cH)$
there exist optimal decompositions (\ref{n2.2}), (\ref{n2.2f}) the
length of which does not exceed
$(\dim \cH)^2 + 1$.

There is a minimal and a maximal roof extension of $g$.

\subsection{Examples}
Roof extensions exist in abundance. To see this and also the
difficulties, we may get into, let us consider a few
examples showing some typical constructions. We remain within the
state space $\Omega(\cH)$, $\dim \cH = d$ and we start with $d =
2$.

\subsubsection{Example 2.1: A Bloch ball construction}
Seen from convex analysis, the space $\Omega$ of all 1-qubit
density operators is a 3-dimensional ball,
the {\em Bloch ball}. $\Omega^{\rm pure}$ is
the {\em Bloch sphere,} the surface of the Bloch ball.

Now assume there is a function $g$ on the Bloch sphere and we like
to find roof extensions $G$ of $g$. Particular nice ones are gained
as following: We take a bundle of straight lines such that every
point $\omega$ of the Bloch ball is coincident with exactly one
line, say $L_{\omega}$, of the bundle. If $\omega$ is not pure, then
$L_{\omega}$ hits the Bloch sphere at exactly two points, say
$\pi_1$ and $\pi_2$. Hence, $\omega$ is a convex combination of
them. Now we define
\begin{equation} \label{e1.0}
G(\omega) = p g(\pi_1) + (1-p)g(\pi_2) \, \hbox{ if } \,
\omega = p \pi_1 + (1-p) \pi_2 \;
\end{equation}

Because there is just one line from our bundle going through
$\omega$, we get a well-defined roof.

Let us now specify our example by choosing a bundle of parallel
lines. It belongs exactly one main axis of the Bloch ball to the
bundle. We may assume that it is the $x_3$-axis with respect to the
Bloch coordinates $x_1, x_2, x_3$ of a general Pauli representation
\begin{equation} \label{e1.1}
\omega = \frac{1}{2}
(\1 + x_1 \sigma_1 + x_2 \sigma_2 + x_3 \sigma_3) \;
\end{equation}

The line $L_{\omega}$ is now fixed by the values $x_1$
and $x_2$, letting the third Bloch coordinate arbitrary.
$L_{\omega}$ crosses the Bloch Sphere at the pure states
$\pi_{\pm}$ with the Bloch coordinates $x_1$, $x_2$,
$y_3 = \pm (1 - x_1^2 - x_2^2)^{1/2}$. Hence
\begin{equation} \label{e1.2}
\pi_{\pm} = \frac{1}{2} (\1 + x_1 \sigma_1 + x_2 \sigma_2 \pm
[1 - x_1^2 - x_2^2]^{1/2}  \sigma_3), \quad
\omega = p \pi_+ + (1-p) \pi_-
\end{equation}
with $0 \leq p \leq 1$. Expressing now
$g$ in terms of Bloch coordinates,
we obtain the roof
\begin{equation} \label{e1.3}
G(\omega) = G(x_1, x_2) = p
g(x_1, x_2, + \sqrt{1 - x_1^2 - x_2^2}) + (1-p)
g(x_1, x_2, - \sqrt{1 - x_1^2 - x_2^2})
\end{equation}

The example is further specified by choosing for $g$
the Shannon entropy of the diagonal elements of $\omega$.
Then $G$ becomes the roof
\begin{equation} \label{e1.7}
E_T(\omega) = H(\frac{1 + \sqrt{1 - x_1^2 - x_2^2}}{2} \; ,
\frac{1 - \sqrt{1 - x_1^2 - x_2^2}}{2} ) \;
\end{equation}

$T(\omega)$ denotes the diagonal part of $\omega$ and
$H$ the Shannon entropy of the diagonal elements.
That this is the solution for the entanglement of the diagonal
channel, has been suggested by Levitin, \cite{Le94}
and Thirring \cite{Th94}. (\ref{e1.7})
provides a roof extension as described above. Indeed,
it is a flat roof depending \linebreak on $1-x_1^2-x_2^2$, hence
a function of the Euclidean
distance $\sqrt{x_1^2 + x_2^2}$ of $\omega$ from the
$x_3$-axis. It is
\begin{equation} \label{e1.8}
\langle 0| \omega |1\rangle = \frac{x_1 - i x_2}{2},
\quad
x_1^2 + x_2^2 = 4 |\,\langle 0, \omega \, |1 \rangle \, |^2
\;
\end{equation}
Therefore, the concurrence of the diagonal map for qubits
can be written
\begin{equation} \label{e1.9}
C_T(\omega) = |\, \langle 0 | \omega  |1 \rangle \, | \;
\end{equation}

It is easily to be seen that $C$ is a flat convex roof. It possesses
quite large domains where it is just affine: Cut the Bloch ball with
a plane containing the $x_3$-axis. We get a disk, cut by the
$x_3$-axis into two half-disks. The concurrence is affine on every
such half-disk.

Indeed, if $T$ is a 1-qubit channel with two different pure fix
points, one can observe a similar phenomenon. (The disks will be cut
by the axis through the fix points.)

\subsubsection{Example 2.2}
There is a constructions, similar to the previous
one: Given any function $f(x_3)$  on the $x_3$-axis. We
construct a roof by
\begin{equation} \label{e1.10}
A(\omega) = f(x_3), \quad x_3 = \langle 0| \omega |0\rangle
 - \langle 1| \omega |1\rangle \;
\end{equation}

More generally, given a bundle of planes so that every point of the
Bloch ball is coincident with one and only one of these planes. A
function, constant on planes, is a roof. This can be achieved by
choosing a function $f$ on the $x_3$-axis and attaching to every
plane the value of $f$ at its crossing with this axis.

Similar constructions are possible with higher dimensional balls
(ellipsoids). However, on higher dimensional state spaces things are
essential more complicated due to the subtle structure of the set
of their pure states.

\subsubsection{Example 2.3: Affine functions on the state space}
An affine function on Herm$(\cH)$ is of the form
\begin{equation} \label{aff1}
X \to l(X) := a + \T \, XA
\end{equation}
with an Hermitian operator $A$ and a real number $a$. It is
sometimes useful to write (\ref{aff1}) in the form
\begin{displaymath}
l(X) = \T \, X B , \quad B = A + \frac{a}{d} \1 \;
\end{displaymath}

The function (\ref{aff1}) is trivially a roof: If we have any
extremal decomposition (\ref{n2.2}) one immediately gets $l(\rho) =
\sum p_j l(\pi_j)$. Thus, every extremal decomposition is optimal.

We easily conclude that with $G$ also $G + l$ is a roof.

A bit more tricky is the assertion that every function (\ref{aff1})
is a {\em flat roof} on the state space $\Omega(\cH)$.

We prove this by induction to the dimension of the Hilbert space. We
start with the qubit case. We can choose a basis in $\cH$ such that
$A = a + a' \sigma_3$ and
\begin{displaymath}
\omega = \frac{1}{2}(\1 + x_1 \sigma_1 + x_3 \sigma_3 ) \;
\end{displaymath}

It suffices to prove the assertion for $A = \sigma_3$,
resulting in $l(X) = \T \, \sigma_3 X = x_3$.
With an unimodular number $\epsilon$ we consider the pure states
\begin{displaymath}
\omega_{\pm} = \frac{1}{2}(\1 + x_1 \sigma_1 \pm
\epsilon \sqrt{1 - x_1^2 - x_3^2} \sigma_2 + x_3 \sigma_3 )
\end{displaymath}
so that our affine function is constant at the segment
$p \omega_+ + (1-p) \omega_-$. The segment contains $\omega$
for $p = 1/2$.

Varying $\epsilon$ we see the following: To every affine function
$l$ as in (\ref{aff1}) there is an axis through the Bloch ball such
that $l$ is constant on every plane perpendicular to that axis. We
can say something more: Given $\omega \in \Omega$ and two affine
functions, $l_j$, $j=1,2$, we use the constants $l_j(\omega) = c_j$
to define two planes by $l_j(X) = c_j$. Because the two planes
contain $\omega$, they intersect along a line, say $L_{\omega}$,
containing $\omega$. The intersection of $L_{\omega}$ with the Bloch
sphere provides two pure states which define a flat optimal
decomposition of $\omega$.
\\[12pt]
\noindent {\bf Proposition 2.2}
\vspace{12pt}

Affine functions on $\Omega(\cH)$, $\dim \cH < \infty$, are flat
roofs. Given two affine functions and a density operator $\omega$,
there is a common flat optimal decomposition of $\omega$.

\begin{proof}
The 2-dimensional case has already been proved. Let $l_j(X) =
a_j + \T \, A_j X$, $j=1,2$,  denote two affine functionals and
choose $\omega \in \Omega(\cH_d)$. Assume the assertion is true for
$\dim \cH < d$. If $\omega$ is from the boundary of $\Omega(\cH_d)$
we are done by our induction hypothesis. Otherwise we consider the
linear subspace $\cL$ of Herm$(\cH_d)$ orthogonal to $A_1$ and
$A_2$, \emph{i.e.}, of all  $Y$ satisfying $\T \, A_j Y = 0$,
$j=1,2$. Consider the affine space $\cL_{\omega} = \omega + \cL$. It
contains $\omega$ and it is $l_j(Y) = l_j(\omega)$ for all $Y \in
\cL_{\omega}$. The intersection $K = \Omega \cap \cL_{\omega}$ is
compact and its extremal points belong to the boundary of
$\Omega(\cH_d)$. Therefore, we \linebreak get a decomposition
\begin{displaymath}
\omega = \sum p_j \omega_j, \quad {\rm rank}(\omega_j) < d
\end{displaymath}
and, furthermore, $l_j(\omega) = l_j(\omega_j)$ because of
$\omega_j \in K$. However, the support of any $\omega_j$
is of dimension less than $d$. Therefore, there are
extremal decompositions
\begin{displaymath}
\omega_j = \sum_k p_{jk} \pi_{jk}, \quad l_j(\omega_j) =
l(\pi_{jk})
\end{displaymath}
for all $j$ and all $k$ with pure states $\pi_{jk}$. Thus
\begin{displaymath}
\omega = \sum_j p_j \sum_k p_{jk} \pi_{jk}
\end{displaymath}
is a flat optimal decomposition of $\omega$ for $l_1$ as
well for $l_2$. Thus the proposition is proved.
\end{proof}
The following is a corollary.
\\[12pt]
\noindent {\bf Proposition 2.3}
\vspace{12pt}

Let $f(x_1, x_2)$ be a function defined on the
range of two affine functions, $l_1$ and $l_2$. Then
\begin{equation} \label{aff2}
F(\omega) := f(l_1(\omega), l_2(\omega))
\end{equation}
is a flat roof.

\subsubsection{Example 2.4: An application to the diagonal map}
We like to apply the proposition above to the diagonal map
\begin{equation} \label{diag1}
X \to D_n(X) = {\rm diag}(X)
\end{equation}
which is a channel on $\Omega(\cH)$. diag$(X)$ denotes the
diagonal part of $X$ obtained by replacing all off-diagonal elements
of $X$ by zeros. Proposition 2.3 proves that the von Neumann entropy
\begin{equation} \label{diag2}
\omega \to S(D_n(\omega)) = \sum \eta(\langle j| \omega |j\rangle)
\end{equation}
is a flat roof for $n= 2, 3$. For $n=3$ one replaces the
third diagonal element, $x_{33}$, by $1 - x_{11} - x_{22}$
to see that proposition 2.3 suffices to verify the assertion.

We like to prove the flatness of (\ref{diag2}) for all dimensions.
This will be done in example 3.2 later on.

\subsubsection{Example 3}
A general way to obtain roof extensions in state spaces is presented
next. It allows for further modifications, but it is difficult to be
controlled explicitly.

Let $g$ be a continuous real function on $\Omega^{\rm pure}$
and $\omega \in \Omega$. Every basis
$|\psi_1\rangle, \dots, |\psi_d\rangle$ gives rise
to a decomposition
\begin{equation} \label{n2.b1}
\omega = \sum \sqrt{\omega} |\psi_j \rangle\langle \psi_j|
\sqrt{\omega}
\end{equation}
of positive rank one operators. After normalization of the rank one
operators we get extremal convex decomposition of $\omega$,
\begin{equation} \label{n2.b2}
\omega = \sum_{p_j \neq 0} p_j
\frac{\sqrt{\omega} |\psi_j \rangle\langle \psi_j|\sqrt{\omega}
}{\langle \psi_j | \omega | \psi_j \rangle} , \quad
p_j = \langle \psi_j | \omega | \psi_j \rangle \,
\end{equation}

In the following definitions we vary over all bases.
\begin{equation} \label{n2.b4}
G_1(\omega) = \min_{\rm bases} \sum_{p_j \neq 0}
p_j g(\frac{\sqrt{\omega} |\psi_j \rangle\langle \psi_j|
\sqrt{\omega}}{\langle \psi_j | \omega | \psi_j \rangle} )
\end{equation}
where $p_j$ is as in (\ref{n2.b2}). Because continuity of
$g$ is assumed, and the set of bases is compact, the minimum
in (\ref{n2.b4}) will be attained. Thus, $G_1$ is a
roof extension of $g$.

One gets another roof extension by
\begin{equation} \label{n2.b3}
G_2(\omega) = \max_{\rm bases} \sum_{p_j \neq 0}
p_j g(\frac{\sqrt{\omega} |\psi_j \rangle\langle \psi_j|
\sqrt{\omega}}{\langle \psi_j | \omega | \psi_j \rangle} ) \;
\end{equation}

Remark: One cannot guaranty the roof property without continuity of $g$.

(\ref{n2.b3}) and (\ref{n2.b4}) are global optimization problems for
which there are no explicit expressions known in most cases. There
are, however, algorithms to approximate them numerically.

\section{Roofs and Convexity}
We start with some definitions and elementary, mostly well-known
statements around maximal convex and minimal concave extensions. The
second subsection is concerned with sufficient conditions for the
existence of convex and concave roofs and their main properties.

\subsection{Convex and Concave Extensions}
Let $\Omega$ be a compact convex set and $g$ a real function
defined on the set $\Omega^{\rm pure}$ of its extremal points.
\\[12pt]
\noindent {\bf Definition 3.1: Convex (concave) extensions}
\vspace{12pt}

A convex function $G$ is called a {\em convex (respectively concave)
extension of $g$} if $G$ is convex (respectively concave) and
coincides on $\Omega^{\rm pure}$ with $g$.

Between any convex, concave or roof extension of a function $g$
the inequalities
\begin{equation} \label{n3.3}
G^{\rm convex} \leq G^{\rm roof} \leq G^{\rm concave}
\end{equation}
are valid. Indeed, with an optimal decomposition
(\ref{n2.2}) for $G^{\rm roof}$ we get
\begin{displaymath}
G^{\rm roof}(\omega) =  \sum p_j G^{\rm roof}(\pi_j)
= \sum p_j G^{\rm convex}(\pi_j)
\end{displaymath}
because the extensions coincide on $\Omega^{\rm pure}$.
The right sum cannot be smaller than
$G^{\rm convex}(\omega)$ by convexity.
Similarly one argues in the concave case.

The proof of (\ref{n3.3}) is valid pointwise, leading to
\\[12pt]
\noindent {\bf Proposition 3.1}
\vspace{12pt}

Let $G^{\rm convex}$ be a convex, $G^{\rm concave}$ a concave, and
$G$ any extension of $g$. If $\omega$ is a roof point of $G$, then
\begin{equation} \label{n3.3a}
G^{\rm convex}(\omega) \leq G(\omega)
\leq G^{\rm concave}(\omega)
\end{equation}

Let $G^{\rm convex}$ and $G^{\rm concave}$ be as in the above
proposition and assume in addition equality in (\ref{n3.3a}),
$G^{\rm convex}(\omega) = G^{\rm concave}(\omega)$. For all convex
combination
\begin{displaymath}
\omega = \sum p_j \omega_j, \quad \omega_j \in \Omega
 , \quad p_j > 0 , \qquad (*)
\end{displaymath}
we conclude
\begin{equation} \label{ng1}
\sum p_j G^{\rm convex}(\omega_j) \geq G^{\rm convex}(\omega) =
G^{\rm concave}(\omega) \geq \sum p_j G^{\rm concave}(\omega_j)
\end{equation}
because of (\ref{n3.3}) the conclusion is
$G^{\rm convex}(\omega_j) = G^{\rm concave}(\omega_j)$
for all $j$.

To express this finding we need some standard terminology. A subset
$K \subset \Omega$ is called a {\em face} of $\Omega$ if from
$\omega \in K$ and (*)it necessarily follows $\omega_j \in K$ for all $j$.
Faces are convex subsets of $\Omega$. If not only $\Omega$ but also
$\Omega^{\rm pure}$ is compact then faces are compact and any face
$K$ is convexly generated by
$\Omega^{\rm pure} \cap K$.\\
The intersection of faces is either empty or a face again. The
smallest face containing $\omega$ will be called $\omega${\em -face}
of $\Omega$ and denoted by face$_{\omega}[\Omega]$. If $K$ is a face
and $\omega$ not a  point of the boundary of $K$, then $K$ is the
$\omega${\em -face} of $\Omega$.

Now we express the finding above by
\newpage
\noindent {\bf Proposition 3.2}
\vspace{12pt}

Coincide a convex and a concave extension of $g$ at $\omega \in
\Omega$, then they coincide on the $\omega$-face of $\Omega$.

The least upper bound of a set of convex functions is convex again.
Hence, given $g$ on $\Omega^{\rm pure}$, there is a unique largest
convex extension of $g$ from $\Omega^{\rm pure}$ to $\Omega$.
Similarly there exists a smallest concave extension of $g$. It is
convenient to introduce an extra notation for these extensions:
\\[12pt]
\noindent {\bf Definition 3.2: $g^{\cap}$} and $g^{\cup}$
\vspace{12pt}

Let $g$ be a real function on $\Omega^{\rm pure}$. We denote by
$g^{\cup}$ the largest convex and by $g^{\cap}$ the smallest concave
extension of $g$ to $\Omega$,
\begin{eqnarray}
 g^{\cup} &=& \, \hbox{ largest convex extension of $g$
 from $\Omega^{\rm pure}$ to $\Omega$ ,}
\nonumber \\
 g^{\cap} &=& \, \hbox{ smallest concave extension of $g$
 from $\Omega^{\rm pure}$ to $\Omega$ .}
\nonumber
\end{eqnarray}

We also write $G = G^{\cup}$ (or $G = G^{\cap}$)
if $G$ is the largest convex (or the smallest concave)
extension of the restriction of $G$ onto
$\Omega^{\rm pure}$.
\\[12pt]
\noindent {\bf Proposition 3.3}
\vspace{12pt}

Let $G$ be an extension of $g$ and $\omega$ one of its roof
points.
If $G$ is convex, then  $G(\omega) = g^{\cup}(\omega)$.
If $G$ is concave, then $G(\omega) = g^{\cup}(\omega)$.

If a convex (resp.~concave) roof extension of $g$ exists, then it is
unique.
Because $G$ is convex, $G \leq g^{\cup}$. Because $\omega$
is a roof point, (\ref{n3.3a}) asserts $G(\omega) \geq
g^{\cup}(\omega)$.

Is there a convex extension at all for a given $g$ ? If there is one
then there is also a largest one, \emph{i.e.} $g^{\cup}$ exists. The
answer to the question is affirmative and has been given in
\cite{Roc70} by a variational characterization which is well known
in quantum information theory as a recipe to construct entanglement
measures:
\begin{eqnarray}
g^{\cup}(\omega) &=& \inf \sum p_j g(\pi_j)
\label{n3.4a} \\
g^{\cap}(\omega) &=& \sup \sum p_j g(\pi_j) \label{n3.4b}
\end{eqnarray}
where the ``inf'', respectively ``sup'', is running over
all extremal convex decompositions
\begin{equation} \label{n3.5}
\omega = \sum p_j \pi_j, \quad \pi_j \in \Omega^{\rm pure}
\end{equation}
of $\omega$.

Indeed, if $G$ is an extension of $g$ which is convex, the right
side of (\ref{n3.4a}) must be always larger than $G(\omega)$. On the
other hand, given $\omega_1$ and $\omega_2$, one can find
decompositions (\ref{n3.5}) for them differing an arbitrary small
amount $\epsilon > o$ from $g^{\cup}(\omega_1)$ respectively
$g^{\cup}(\omega_2)$. They may be composed from the pure states
$\pi_{i,j}$ and probabilities $p_{i,j}$, $i = 1,2$. Then
\begin{displaymath}
2 \epsilon + p g^{\cup}(\omega_1) + (1-p) g^{\cup}(\omega_2)
\geq \sum p p_{1,j} g(\pi_{1,j}) + \sum (1-p) g(\pi_{2,k})
\end{displaymath}
and this is not smaller than
$g^{\cup}(p \omega_1 + (1-p) \omega_2)$. Because $\epsilon$
can be arbitrary near to zero, $g^{\cup}$ is convex.
The concave case can be settled by a similar reasoning or by
\begin{equation} \label{n3.6}
- g^{\cup} = (-g)^{\cap}, \quad - g^{\cap} = (-g)^{\cup}
\end{equation}
\\[12pt]
\noindent \underline{Remark:} Hulls of functions
\vspace{12pt}

The {\em convex hull} of $G$ is the largest convex function which is
smaller than $G$. The {\em concave hull} of $G$ is the smallest
concave function which is larger than $G$. In (\ref{n3.4a}) and
(\ref{n3.4b}) one uses the values of $G$ at the pure states only.
Because the hull construction must respect values on the whole of
$\Omega$, there are more constraints to be fulfilled.

The expressions (\ref{n3.4a}) and (\ref{n3.4b}) are similarly
structured as those of the convex and the concave hulls of a
function $G$ on $\Omega$. One mimics the proofs and gets
\begin{eqnarray}
{\rm conv}[G](\omega) &=& \inf \sum p_j G(\omega_j)
\label{h3.4a} \\
{\rm conc}[G](\omega) &=& \sup \sum p_j G(\omega_j) \label{h3.4b}
\end{eqnarray}
where, as in (\ref{ng1}), one has to run through
all convex combinations
\begin{displaymath}
\omega = \sum p_j \omega_j, \quad \omega_j \in \Omega
 , \quad p_j > 0 \;
\end{displaymath}

Obviously, conv$[G] \leq G^{\cup}$ and conc$[G] \geq G^{\cap}$.
This quite simple reasoning provides also
\\[12pt]
\noindent {\bf Proposition 3.4}
\vspace{12pt}

If $G$ is a concave or a roof extension of $g$ then
$g^{\cup} = {\rm conv}[G]$.

If $G$ is a convex or a roof extension of $g$ then
$g^{\cap} = {\rm conc}[G]$.

As a matter of fact one can do similar hull constructions with any
subset of $\Omega$ which convexly generates $\Omega$. This has been
emphasized in \cite{VW96}.

\subsection{Convex and Concave Roofs}
If there is a convex roof extension of $g$, then it is equal to
$g^{\cup}$. There is a sufficient condition to guaranty the roof
property of $g^{\cup}$ and of $g^{\cap}$.
\\[12pt]
\noindent {\bf Proposition 3.5}
\vspace{12pt}

Let $\Omega$ be a convex set. Assume both, $\Omega$ and
$\Omega^{\rm pure}$, are compact and $g$ continuous on
$\Omega^{\rm pure}$. Then
$g^{\cup}$ and $g^{\cap}$ are roofs. According to proposition 3.3
they are the minimal respectively maximal roof extensions \linebreak of $g$.

Remember that $\Omega^{\rm pure}$ is compact if $\Omega =
\Omega(\cH)$ and $\cH$ is finite dimensional. The requirement of
continuity of $g$ is often satisfied in physically motivated
applications, though not always. A counter example is the Schmidt
number in bipartite quantum systems. In this case it is not known
whether $g^{\cup}$ and $g^{\cap}$ are roofs. Nevertheless. the
assumptions needed are rather weak ones. We met them already in
proposition 2.1.

The proof will be ``constructive'' in a certain sense. It is
arranged to sharpen ``theorem 1'' in \cite{BNU96}: If we know an
optimal decomposition with pure states $\pi_1, \pi_2, \dots$, then
every convex combination of them is optimal. In particular, the
(convex or concave) roof is affine on the convex set generated by
the pure states $\pi_1, \pi_2, \dots$. Restricted to this set,
the graph of $G$ is a piece of an affine space. The whole
graph of $G$ appears as composed of affine pieces \cite{fnote1}.
If one would know a covering of $\Omega$ by these ``convex leaves'',
one could compute $g^{\cup}$ from the values of $g$ at
$\Omega^{\rm pure}$. Things are similar for $g^{\cap}$.

To start proving propositions 3.5 and 2.1 let us repeat the
assumptions. $\Omega$ is a compact convex set in a real linear space
$\cL$ of finite dimension,  the set $\Omega^{\rm pure}$ of all pure
(\emph{i.e.}, extremal) points of $\Omega$ is compact. $g$ is a real
continuous function on $\Omega^{\rm pure}$. The dimension of
$\Omega$ as a set in $\cL$ is denoted by $n$. It is the dimension
of the affine space generated by $\Omega$.\\
{\em Remark:} The space $\Omega(\cH)$ of density operators is
embedded in Herm$(\cH)$. The latter is of dimension $d^2$
if $\dim \cH = d$. The affine space generated by
$\Omega(\cH)$ is the hyperplane of Hermitian operators of
trace one. The dimension of $\Omega(\cH)$ is $n = d^2 -1$.

We enlarge $\cL$ to the linear space $\cL' = \cL \oplus \bbbr$. Its
elements, $X \oplus \lambda$, will be written in
 vector form $\{X, \lambda \}$ with two components,
$X \in \cL$ and $\lambda \in \bbbr$.
We need the set
\begin{equation} \label{32.1}
E = \{ \, \pi, \, g(\pi) \,\} , \quad
\pi \in \Omega^{\rm pure} \;
\end{equation}
$E$ is a compact set by our assumption. Hence its convex hull,
denoted by $\Omega[g]$, is a compact convex set. The set of extremal
points of $\Omega[g]$ is $E$. (If one of the elements of $E$ would
be a convex combination of the others, the same would be true for
the corresponding pure states, contradicting our assumptions.)

Choose $\omega \in \Omega$ and consider in $\cL \oplus \bbbr$ the
straight line consisting of the points $\{\omega, \lambda \}$,
$\lambda \in \bbbr$. The line intersects with $\Omega[g]$ along a
compact segment
\begin{equation} \label{32.2}
\{\omega, \lambda \} \in \Omega[g] \, \Leftrightarrow \,
\lambda_0(\omega) \leq \lambda \leq \lambda_1(\omega) \;
\end{equation}
$\lambda$ satisfies (\ref{32.2}) if and only if there
is an extremal decomposition
\begin{equation} \label{32.3}
\{\omega, \lambda \} =
\sum p_j \{\pi_j, g(\pi_j) \} = \{ \, \omega, \, \sum p_j g(\pi_j) \,\},
\quad \pi_j \in \Omega^{\rm pure} \;
\end{equation}
Therefore,
\begin{equation} \label{32.4}
g^{\cup}(\omega) = \lambda_0(\omega) \leq \lambda \leq \lambda_1(\omega) =
g^{\cap}(\omega)
\end{equation}
and there exist extremal decompositions of $\omega$ with
equality in (\ref{n3.4a}) respectively (\ref{n3.4b}).
Therefore, $g^{\cup}$ and $g^{\cap}$ are roofs and
proposition 3.5 has been proved.

If $G$ is a roof extension of $g$, the point $\{\omega, G(\omega)\}$
is contained in $\Omega[g]$. Hence it can be represented by a convex
combination of elements from $E$. As the dimension of $\Omega[g]$ is
$n+1$, there are, by a theorem of Carath\`eodory, pure convex
decomposition of length $n+2$. This proves proposition 2.1: For
$\Omega = \Omega(\cH)$ it follows $n+2 = (d^2 -1) +2$.

On the other hand, a point $\{\omega, g^{\cup}(\omega)\}$ belongs to
a face of the boundary of $\Omega[g]$. Its dimension cannot exceed
$n$. Thus there are, again by Carath\`eodory, pure decompositions of
length $n+1$. For $\Omega(\cH)$ this  gives an achievable length
$d^2$, an often used fact \cite{fnote2}.
\\[12pt]
\noindent {\bf Definition 3.2: convex leaves}
\vspace{12pt}

Let $G$ be a real function on $\Omega$.
A subset $K \subset \Omega$ is called a {\em convex leaf}
of $G$ if
\vspace{6pt}

(a) K is  compact and convex,

(b) K is the convex hull of $K \cap \Omega^{\rm pure}$,

(c) G is convexly linear on $K$, \emph{i.e.}
\begin{displaymath}
G(\sum p_j \rho_j) = \sum p_j G(\rho_j) \,
\hbox{ if all } \, \rho_j \in K \;
\end{displaymath}

$K$ is called {\em complete,} if all pure states which can
appear in an optimal decomposition of any $\rho \in K$
are contained in $K$.
\\[12pt]
\noindent {\bf Proposition 3.6}
\vspace{12pt}

With the assumptions of proposition 3.5 it holds: For $g^{\cup}$
(respectively $g^{\cap})$ and any $\omega \in \Omega$ there are
complete convex leaves containing $\omega$.

It makes sense to call the set of all complete convex leaves of $G$
the {\em convex foliation} of $G$ or, shortly, the $G$-foliation.
\begin{proof}
The proposition will be proved for
$g^{\cup}$. The case of $g^{\cap}$ is similar.
There is an affine functions $l$ such that
\begin{equation} \label{leaf1}
l \leq g^{\cup} , \quad l(\omega) = g^{\cup}(\omega)
\end{equation}
We define a subset $K$ of $\Omega$ by
\begin{equation} \label{leaf2}
K = \{ \rho \, | \, l(\rho) = g^{\cup}(\rho) \ \}
\end{equation}
\end{proof}
\noindent {\bf Proposition 3.6.a:  \, (\ref{leaf2}) is a
complete $g^{\cup}$-leaf.}
\vspace{12pt}

Clearly, $\omega \in K$. Let us choose $\rho \in K$.
For an optimal pure decomposition, $\rho = \sum p_j \pi_j$,
one obtains
\begin{displaymath}
\sum p_j g^{\cup}(\pi_j) =
g^{\cup}(\rho) = l(\rho) = \sum p_j l(\pi_j)
\end{displaymath}

However, $l(\pi_j) \leq g^{\cup}(\pi_j)$ by assumption. By the
equation above, all these inequalities must be equalities. Hence,
{\em the pure composers $\pi_j$ of every optimal decomposition of
any $\rho \in K$ are contained in $K$,} \emph{i.e.}, $K$ is
complete. Now choose another $\rho' \in K$ and let $\rho' = \sum
p_k' \pi_k'$ be an optimal decomposition. For $0 < p < 1$ we get,
applying first our assumption and then convexity of $g^{\cup}$,
\begin{displaymath}
l(p \rho + (1-p) \rho') \leq g^{\cup}(p \rho + (1-p) \rho')
\leq p g^{\cup}(\rho) + (1-p)g^{\cup}(\rho')
\end{displaymath}

By the assumption the right hand side can be written
\begin{displaymath}
p l(\rho) + (1-p)l(\rho') = l(p \rho + (1-p) \rho')
\end{displaymath}
and is equal to the left expression. Hence, equality must hold
and $K$ must be a convex set. We have seen already that every
$\rho \in K$ can be represented by a convex combination of
elements from $K \cap \Omega^{\rm pure}$. Because $g$ is
continuous, the set of all $\pi$ satisfying $g(\pi) = l(\pi)$
is compact. Hence, $K$ is compact. Indeed, it is convexly
generated by a compact set. $\Box$

We repeat a further standard notation. An element $\rho$ of a convex
set $K$ is called $K${\em -inner} or ``convexly inner'' if for any
$\nu \in K$, and for small enough positive $s$, it
follows $(1+s) \rho - s \nu \in K$.
Geometrically, the line segment from $\nu$ to $\rho$
can be prolonged a bit without leaving $K$. There is also a
topological characterization: A $K$-inner point is an inner point of
$K$ with respect to the affine space generated by $K$. (Example: The
invertible density operators are the convexly inner points of
$\Omega(\cH)$.)

The intersection of complete convex leaves is either empty or it is
a complete convex leaf. Hence there is a minimal complete convex
leaf containing a given $\omega \in \Omega$, the {\em $\omega$-leaf}
of $g^{\cup}$. It is convexly generated by all those $\pi \in
\Omega^{\rm pure}$ which
can appear in an optimal decomposition of $\omega$.

The $\omega$-leaf is the largest convex leaf containing
$\omega$ as convexly inner point.

Now let $K_1$ be the $\omega_1$-leaf and $K_2$ that of
$\omega_2$. If $\omega_1$ is a convexly inner point of
$K_2$, then $K{_1} \subset K_2$.
In particular, $K_1 = K_2$ if and only if they
contain a point which is commonly inner. Let us draw
a corollary:

If $K_2$ is properly larger than $K_1$, the convex dimension of
$K_2$ must be strictly larger than that of $K_1$.

A chain $K_1 \subset K_2 \subset \dots$, consisting of different
complete $g^{\cup}$-leaves, cannot contain more than $n+1$ members.
As above, $n$ denotes the convex dimension of $\Omega$. The maximal
number of different leaves in any chain is called the {\em depth} of
the $g^{\cup}$-foliation. As said above, the $g^{\cup}$-foliation
consists of all complete $g^{\cup}$-leaves.

In the case $\Omega = \Omega(\cH)$, the depth if bounded by $d^2$.
If the roof is an affine function, see example 2.3, the
faces of $\Omega$ are exactly its leaves and the bound is reached.

Let us look again to the setting above in more geometric
terms. We shall see that the $\omega$-leaves of $g^{\cup}$
and $g^{\cap}$ correspond uniquely to the faces of
$\Omega[g]$.

The triple $\{\Omega[g], \Omega, \Pi\}$ is a
{\em fiber bundle} with bundle space $\Omega[g]$,
base space $\Omega$, and projection
\begin{equation} \label{32.5}
\Pi \, : \quad \{ \omega, \lambda \} \to \omega \;
\end{equation}
In this scheme a roof $G$ becomes a {\em cross section},
say $s_G$, by setting
\begin{equation} \label{32.6}
\omega \to s_G(\omega) = \{ \omega, G(\omega)\} \in \Omega[g] ÿ\;
\end{equation}
we get $\Pi(s_G(\omega)) = \omega$, which is necessary
for a bundle structure.

The boundary, $\partial \Omega[g]$, of $\Omega[g]$ is the
union of three disjunct sets:
\begin{eqnarray}
\partial^{0} \Omega[g] &=&
\{\{\omega, g^{\cup}(\omega) \} \,| \,
g^{\cup}(\omega) = g^{\cap}(\omega) \}
\label{32.7a} \\
\partial^{-} \Omega[g] &=&
\{\{\omega, g^{\cup}(\omega) \} \,| \,
g^{\cup}(\omega) \neq g^{\cap}(\omega) \}
\label{32.7b} \\
\partial^{+} \Omega[g] &=&
\{\{\omega, g^{\cap}(\omega) \} \,| \,
g^{\cup}(\omega) \neq g^{\cap}(\omega) \}
\label{32.7c}
\end{eqnarray}
The cross section (\ref{32.6}) with $G = g^{\cup}$ maps
$\Omega$ onto
$\partial^{0} \Omega[g] \cup \partial^{-} \Omega[g]$
while the cross section with $G = g^{\cap}$ maps
the base space onto
$\partial^{0} \Omega[g] \cup \partial^{+} \Omega[g]$.

The fibres degenerate to a point at the boundary part (\ref{32.7a})
and can be identified with a subset of $\Omega$. By proposition 3.2
that subset consists of the pure point and possibly of some faces at
which $g^{\cup} = g^{\cap}$, and all roof extensions of $g$ coincide
and are affine.

Let us consider a face $\tilde K$ contained in the
``lower'' part $\partial^{-} \Omega[g]$ of the
boundary (\ref{32.7b}).
$\tilde K$ is convex by definition and compact because of the
compactness of $\Omega[g]^{\rm pure}$.
Therefore, the projection $K$ of $\tilde K$ to $\Omega$,
$\Pi \, \tilde K = K$, is convex and compact, see (\ref{32.5}).
We use the supposed face property:
 $\tilde \omega \in \tilde K$ implies
\begin{equation} \label{32.8}
\tilde \omega = \sum p_j \tilde \omega_j \, \Rightarrow \,
\tilde \omega_j \in \tilde K
\end{equation}
if all $p_j > 0$. Writing this out in the manner
$\tilde \omega = \{\omega, g^{\cup}(\omega)\}$, and so on,
we arrive at
\begin{equation} \label{32.9}
\{\omega, g^{\cup}(\omega)\} = \sum p_j
\{\omega_j, g^{\cup}(\omega_j)\} =
\{\omega, \sum p_j g^{\cup}(\omega_j)\} \;
\end{equation}

Now we can state
\\[12pt]
\noindent {\bf Proposition 3.7}
\vspace{12pt}

There is a one-to-one correspondence between the faces of
$\Omega[g]$ contained in $\partial^{0} \Omega[g] \cup \partial^{-}
\Omega[g]$ and the complete convex leaves of $g^{\cup}$. The cross
section $s_G$, with $G = g^{\cup}$, maps complete convex leaves of
$g^{\cup}$ onto faces of $\Omega[g]$. The bundle projection $\Pi$
returns them back to
$\Omega$.\\
For the concave roof things are similar.

\subsection{Illustrating Examples}

\subsubsection{Example 3.1: Minimum and maximum of $g$}
On $\Omega^{\rm pure}$ let $g_{\min}$ be the minimum
of $g$ and $g_{\max}$ its maximum.
The convex hull of the set
\begin{equation} \label{3e.1}
\{ \pi \in \Omega^{\rm pure} \, | \, g(\pi) = g_{\min} \}
\end{equation}
is a complete convex leaf of $g^{\cup}$.
The convex hull of the set
\begin{equation} \label{3e.2}
\{ \pi \in \Omega^{\rm pure} \, | \, g(\pi) = g_{\rm max} \}
\end{equation}
is a complete convex leaf of $g^{\cap}$.

Let us again consider the diagonal map $D(X) = {\rm diag}(X)$ as in
(\ref{diag1}) and its von Neumann entropy $S(D(.))$, see
(\ref{diag2}).  $g(\pi) = S(D(\pi))$ is the output entropy of the
pure state $\pi$. The Hilbert space dimension is denoted by $d$. As
well known, the minimum output entropy is zero and the maximal
\linebreak one $\log d$.

Things become more refined by restricting the channel onto a face of
$\Omega$. As an example we take a short look at the
$(d-1)$-dimensional subspace $\cH_0$ which is orthogonal to the
vector $|\varphi\rangle = d^{-1/2} \sum |j\rangle$. $\cH_0$ consists
of vectors $\sum a_j|j\rangle$ such that $\sum a_j = 0$. $\cH_0$
supports some pure states satisfying $D(\pi) = d^{-1} \1$ and the
maximal output entropy is $\log d$ again.

There is a reasonable conjecture, saying  that the
{\em minimal output entropy is independent of $d$ and
equal to $\log 2$.} There are
$d(d-1)/2$ pure states $\pi_{jk}$, $j< k$. The matrix
elements $a_{nm}$ of $\pi_{jk}$
are $1/2$ for $n=m=j$ and $n=m=k$. They are
$-1/2$ for $n=j, m=k$ and $m=j, n=k$, and all other entries
are zeros. Hence it is evident that
$S({\rm diag}(\pi_{jk})) = \log 2$ and, therefore,
\begin{displaymath}
E_D(\rho) \leq \log2, \quad \rho =
\frac{1}{d-1} (\1 - |\varphi \rangle\langle \varphi|)
\end{displaymath}
because we can represent $\rho$ by a convex combination of the pure
states $\pi_{jk}$. The conjecture asserts that the decomposition is
an optimal one. The conjecture rests on the fact that for no other
state than $\pi_{jk}$, supported by $\cH_0$, the output of the
diagonal map is of rank two. Then one applies a theorem of Michelson
and Jozsa, see appendix of \cite{MJ03}, reducing in the case at hand
the minimization of the entropy to that of minimizing the second
elementary symmetric function or, equivalently, to the minimization
of the concurrence. For $d=2$ this is trivial, for $d=3$ it can be
done, for $d>3$ it is yet a conjecture.

Under the assumption, the conjecture is true, the convex set
generated by the $\pi_{jk}$ is a complete leaf. The minimal length
of an optimal decomposition of $\rho$ is equal to $d(d-1)/2$.

\subsubsection{Example 3.2: Again the diagonal map}
We return to the example 2.4 to prove the flatness of the
Entropy (\ref{diag2}) of the diagonal map (\ref{diag1}).
Hence, since we know from von Neumann's work the
concavity of $\omega \to S(D(\omega))$, flatness of
 (\ref{diag2}) proves the diagonal map a flat concave
roof for all dimensions $d$ of $\cH$. In other words, for
\begin{equation} \label{3e.3}
g(\pi) = S({\rm diag}(\pi)), \quad \pi \in \Omega(\cH)
\end{equation}
it follows for any density operator $\omega$
\begin{equation} \label{3e.4}
g^{\cap}(\omega) = S({\rm diag}(\omega)) \;
\end{equation}
\noindent {\bf Proposition 3.8}
\vspace{12pt}

Choose $\omega \in \Omega(\cH)$. The set $K_{\omega}$ of
all states $\rho$ with ${\rm diag}(\rho) = {\rm diag}(\omega)$
is a complete convex leaf of (\ref{3e.4}). $g^{\cap}$
is a flat concave roof.

The set $K_{\omega}$ is compact, convex, and $S(D(.))$ is constant
on it.  Transversal to the sets $K_{\omega}$ the function $S(D(.))$
is strictly concave. This excludes that two density operators with
different diagonals can belong to a convex leaf. Hence, the sets
$K_{\omega}$ are convex leaves.

\section{Wootters' Method}
We are going to describe the fundamental idea in \cite{Woo97}, see
also \cite{HilWoo97}, and its generalizations \cite{Uh00c}. After a
short introduction to anti-linearity, which is on the heart of the
method, we present slightly simplified proofs for a class of convex
and concave roofs. For reasons of uniqueness we sometimes write
$\langle .,. \rangle$ for the scalar product instead of Dirac's
$\langle .|. \rangle$.

The use of anti-linearity \cite{fnote3}
goes back, at least in physics,
to Wigner. He applied it to the time reversal symmetry
\cite{Wig32} and he discovered the structure of anti-unitary
operators, \cite{Wi60}.
A highlight in the further development of this line of
thinking is the proof of the CPT-theorem
within Wightman's axiomatic quantum field theory.

\subsection{Anti-Linearity in Short}
We start with some elementary remarks. An anti-linear
operator, say $\vartheta$, obeys the rule
\begin{equation} \label{rule1}
\vartheta ( \, a_1 |\phi_1\rangle + a_2 |\phi_2\rangle \, ) =
a_1^* \vartheta |\phi_1\rangle + a_2^* \vartheta |\phi_2\rangle
\end{equation}
An important fact follows immediately: Because of
$c \vartheta = \vartheta c^*$ the {\rm eigenvalues of
$\vartheta$
form a set of circles.} Indeed, if $|x\rangle$ is an
eigenvector of $\vartheta$ with eigenvalue $a$, then
$\epsilon |x\rangle$, $|\epsilon|=1$,
is an eigenvector with eigenvalue $\epsilon^* a$.
Consequently, most of the unitary invariants of linear
operators are undefined for anti-linear ones. The
trace, for example, does not exist for anti-linear operators.

The Hermitian adjoint $\vartheta^{\dag}$ of an anti-linear
operator $\vartheta$ is defined by
\begin{equation} \label{rule2}
\langle \phi_1 , \vartheta^{\dag} \,  \phi_2 \rangle  =
\langle \phi_2 , \vartheta \, \phi_1 \rangle
\end{equation}
There is to set a caution mark: Do not apply an anti-linear operator
to a bra in the usual Dirac manner! By (\ref{rule2}) one may get
absurd results.

A useful class of anti-linear operators are the Hermitian ones. By
(\ref{rule2}) any matrix representation must result in a complex
symmetric matrix. About symmetric matrices see \cite{Horn90}. It
follows that the Hermitian anti-linear operators constitute a
complex linear space of dimension $d(d+1)/2$ if $\dim \cH = d$.

An anti-unitary, $V$, is an anti-linear operator which is unitary,
\emph{i.e.}, satisfies $V^{\dag} = V^{-1}$. A conjugation, $\Theta$,
is an anti-unitary which is Hermitian. It implies $\Theta^2 = \1$.
In accordance with what has been said about eigenvalues, one can
find an orthogonal basis $|\phi_1\rangle, \dots$ such that $\Theta
|\phi_j\rangle = \epsilon_j |\phi_j \rangle$ with arbitrarily chosen
unimodular numbers $\epsilon_j$.

A conjugation $\Theta$ distinguishes a {\em real} Hilbert
subspace $\cH_{\Theta}$ of $\cH$ consisting of all
$\Theta$-real vectors, $\Theta |\psi\rangle = |\psi\rangle$.

There is a polar decomposition, $\vartheta = V |\vartheta|$, for any
anti-linear operator $\vartheta$.  $|\vartheta|$ denotes the
positive root $(\vartheta^{\dag} \vartheta)^{1/2}$ and $V$ is an
anti-unitary operator. The proof is similar to the linear case
\cite{fnote4}.

Now we turn to the case of an anti-linear Hermitian operator
$\vartheta = \vartheta^{\dag}$. It commutes with the positive
(linear!) operator $\vartheta^2$ and, therefore, with $|\vartheta| =
(\vartheta^2)^{1/2}$. With a non-singular $\vartheta$ we perform
$\Theta = \vartheta^{-1} |\vartheta|$ the square of which is $\1$.
As it is Hermitian too, it is a conjugation. We conclude
\begin{equation} \label{Polardeco1}
\vartheta = \Theta |\vartheta| = |\vartheta| \Theta, \quad
\Theta  = \Theta^{\dag} = \Theta^{-1}       \;
\end{equation}

By continuity, or by a more detailed analysis, (\ref{Polardeco1})
can be verified for all anti-linear Hermitian $\vartheta$.

\subsection{Building Roofs with an Anti-Linear Hermitian
$\vartheta$} Let $\vartheta$ be Hermitian and anti-linear on a
Hilbert space $\cH$ of dimension $d$. This setting provides
\linebreak a function
\begin{equation} \label{aroof0}
g(\pi) = | \langle \psi, \vartheta \, \psi\rangle |, \quad
\pi = |\psi \rangle\langle \psi|
\end{equation}
on the pure states of $\Omega(\cH)$. Now $g^{\cap}$ and $g^{\cup}$
are well defined and we shall prove:
\\[12pt]
\noindent {\bf Proposition 4.1}
\vspace{12pt}

Let $g$ be as in (\ref{aroof0}).
If $\{\lambda_1 \geq \lambda_2 \geq \dots \}$ denote the eigenvalues
of $| \sqrt{\omega} \vartheta \sqrt{\omega} |$, then
\begin{eqnarray}
g^{\cup}(\omega) &=&
\max \{0, \; \lambda_1 - \sum_{j>1} \lambda_j \}
, \label{aroofth1a} \\
g^{\cap}(\omega) &=& \sum \lambda_j
\label{aroofth1b} \;
\end{eqnarray}
$g^{\cup}$ and $g^{\cap}$ are flat roofs.

At first we simplify the assertion by starting with
\begin{equation} \label{aroofth2}
\vartheta_{\omega} := \sqrt{\omega} \vartheta \sqrt{\omega}
, \quad
|\vartheta_{\omega}  | =  \big(   \sqrt{\omega} \,
 \vartheta \omega \vartheta  \, \sqrt{\omega}  \big)^{1/2} \;
\end{equation}

Up to normalization every pure decomposition of $\omega$
can be gained from a decomposition of the unit operator $\1$,
\begin{equation} \label{aroofth3}
\omega = \sum  \sqrt{\omega} \pi_j \sqrt{\omega} ,
 \quad
\1 = \sum \pi_j \;
\end{equation}
(\ref{aroof0}) is 1-homogeneous on the positive rank one
operators. Comparing (\ref{aroofth3}) with (\ref{n3.4a})
and (\ref{n3.4b}), it can be seen that
\begin{equation} \label{aroofth4}
g^{\cup}(\omega) = \inf \sum
| \langle \psi_j , \vartheta_{\omega} \psi_j \rangle |
 \quad ,
g^{\cap}(\omega) =  \sup \sum
| \langle \psi_j , \vartheta_{\omega} \psi_j \rangle |
\end{equation}
where we have to run through all rank one decompositions
of $\1$,
\begin{equation} \label{aroofth5}
              \sum | \psi_j \rangle \langle \psi_j | = \1 \;
\end{equation}

Now $\vartheta_{\omega}$ is Hermitian and anti-linear. Therefore
there is to any chosen set of phase factors $\epsilon_1, \epsilon_2,
\dots$ a basis $\varphi_1, \, \varphi_2, \, \dots$ satisfying
\begin{equation} \label{aroofth6}
|\vartheta_{\omega} | \,  |\varphi_j \rangle =
\lambda_j |\varphi_j \rangle  , \quad
 \vartheta_{\omega} \,  |\varphi_j \rangle =
\lambda_j \epsilon_j  |\varphi_j \rangle \;
\end{equation}
and the conjugation $\Theta$ in the
polar decomposition multiplies the j-th basis
vector by $\epsilon_j$.

For the next step we assume the existence of a real $d \times d$
Hadamard matrix. Then we can choose a basis $\{|\chi_i\rangle \}$
fulfilling
\begin{equation} \label{aroofth7}
|\chi_i\rangle = \frac{1}{\sqrt{d}} \sum_j a_{ij}
|\varphi_j\rangle , \quad a_{ij} = \pm 1 \;
\end{equation}
because of the orthogonality and $a_{ki}^2 = 1$ we get for all $k$
\begin{equation} \label{aroofth8}
d \, \langle \chi_k, \vartheta_{\omega} \chi_k\rangle  =
 \sum_{ij} a_{ki} a_{kj}
\langle \varphi_i, \vartheta_{\omega} \varphi_j\rangle
= \sum \epsilon_j \lambda_j
\end{equation}

Therefore, by (\ref{aroof0}), we get
\begin{equation} \label{aroofth9}
g^{\cup}(\omega) \leq | \, \sum \epsilon_j \lambda_j \, |
\leq g^{\cap}(\omega) \;
\end{equation}

By varying the unimodular numbers $\epsilon_j$, which could
be chosen arbitrarily, one arrives at
\begin{equation} \label{aroofth10}
g^{\cup}(\omega) \leq \max \{0, \; \lambda_1 - \sum_{j>1} \lambda_j \}
 , \quad
g^{\cap}(\omega) \geq \sum \lambda_j \quad
\end{equation}

Assuming equality in (\ref{aroofth10}), we see from (\ref{aroofth8})
that $\omega$ is a flat point of $g^{\cap}$ and of $g^{\cup}$.
If $g^{\cup}(\omega) > 0$, then we choose $\epsilon_1 = 1$ and
$\epsilon_j = -1$ for $j > 1$ to obtain from (\ref{aroofth6})
an optimal basis $\{|\chi_k \}$. In the concave case we
set $\epsilon_j = 1$ for all $j$.

Now we are going to prove equality in (\ref{aroofth10}), starting
with $g^{\cup}$. We clearly get an estimation from below in
(\ref{aroofth4}) by
\begin{equation} \label{aroofth11}
\inf | \,\sum \xi_k | \; , ÿ \quad \xi_k =
 \langle \psi_k , \vartheta_{\omega} \psi_k \rangle| \;
\end{equation}
Sandwiching with the eigenbasis (\ref{aroofth6}) it yields
\begin{displaymath}
\sum \xi_j = \sum_{jk} \langle \psi_k, \vartheta_{\omega}
\varphi_j \rangle \, \langle \varphi_j, \psi_k\rangle =
\sum_{jk} \epsilon_j \lambda_j
\langle \psi_k, \varphi_j \rangle \,
\langle \varphi_j, \psi_k \rangle
\end{displaymath}
and this shows, summing first over $k$,
\begin{displaymath}
| \sum_k \xi_k | = | \sum_k \epsilon_j \lambda_j | \;
\end{displaymath}
Its minimum is attained by the largest of the two numbers
$0$ and $\lambda_1 - \lambda_2 - \dots$ and the first
equation in (\ref{aroofth2}) is true.

Now we prove equality in (\ref{aroofth10}) for $g^{\cap}$. the
second equation of (\ref{aroofth4}). It is
\begin{displaymath}
| \langle \psi_k, \vartheta_{\omega} \psi_k \rangle | =
| \langle \psi_k, |\vartheta_{\omega}| \psi_k' \rangle |
 , \quad |\psi_k'\rangle =  \Theta |\psi_k\rangle \;
\end{displaymath}
 We
apply Cauchy's inequality. The result is
\begin{displaymath}
| \langle \psi_k, |\vartheta_{\omega}| \psi_k' \rangle |^2
\leq
 \langle \psi_k, |\vartheta_{\omega}| \psi_k \rangle \,
\langle \psi_k', |\vartheta_{\omega}| \psi_k' \rangle
\end{displaymath}
and, because $\Theta$ is an involution, hence anti-unitary,
we arrive at
\begin{equation} \label{aroofth12}
| \langle \psi_k, \vartheta_{\omega} \psi_k \rangle |
\leq
\langle \psi_k, |\vartheta_{\omega}| \psi_k \rangle \;
\end{equation}
Summing up we get the trace of $|\vartheta_{\omega}|$ which upper
bounds $g^{\cap}(\omega)$.

Up to now the proof of proposition 4.1 rests on particular bases $\{
|\chi_k\rangle \}$. They exist if there is a real $d \times d$
Hadamard matrix, $d = \dim \cH$. To overcome the restriction we go
to a larger Hilbert space, $\cH \oplus \cH_0$, for which the
proposition has been proved. Then we restrict to the face of density
operators supported by the original Hilbert space $\cH$. Because a
(flat) roof remains a (flat) roof if restricted to a face, the proof
will become complete.

To do so, we choose $d' = \dim \cH \oplus \cH_0$ sufficiently large
and extend $\vartheta$ to $\vartheta'$ by requiring $\vartheta'
|\psi_0\rangle = 0$ for all $|\psi_0\rangle \in \cH_0$. We then
choose any conjugation $\Theta_0$ on $\cH_0$ and use $\Theta' =
\Theta \oplus \Theta_0$. Now, if there is a real $d' \times d'$
Hadamard matrix, we are done.

Hadamard matrices exist in dimensions $d' = 2^m$. This suffices for
the proof.
\\[12pt]
\noindent {\bf Proposition 4.2}
\vspace{12pt}

Let $g(\pi) = |\langle \psi, \vartheta \psi \rangle|$. Then
$g^{\cup}$ and $g^{\cap}$ allow for flat optimal decompositions of
length $d'$ where \linebreak $d \leq d' = 2^m$. More generally, if
$d \leq d'$ and there is a real $d' \times d'$ Hadamard matrix, then
there are flat optimal decompositions of length $d'$.

\subsection{Cases of Application}
Indeed, the question is now: How to find a suitable anti-linear
Hermitian operator $\vartheta$ to calculate concurrence, tangle, and
entanglement entropy (as particular cases of entanglement of
formation) \linebreak in $2 \times n$ systems. Clearly, this can be
fully successful for flat roofs only.

Let $T$ be a trace preserving positive map from the states
\begin{displaymath}
\omega \in \Omega^d := \Omega(\cH), \quad \dim \cH = d \;
\end{displaymath}
into the 1-qubit state space $\Omega(\cH_2)$. Because the trace of
the output, $\T \, T(\omega)$ is one, $T(\omega)$ is characterized,
up to a unitary transformation, by one variable. It is common to use
$4 \det T(\omega)$ or its square root to be this variable. Let us
abbreviate the convex roofs on $\Omega^d$, playing a role below. By
\begin{eqnarray}
C_T &=& 2 (\sqrt{ \det T})^{\cup}, \quad
(\det T)(\omega) = \det T(\omega) \;
\label{4q.1a} \\
\tau_T &=& 4 (\det T)^{\cup} \quad \;
\label{4q.1b} \\
E_T &=& (S_T)^{\cup}, \quad S_T(\omega) = S( T(\omega) )
\label{4q.1c}
\end{eqnarray}
(\ref{4q.1a}) is the concurrence, (\ref{4q.1b}) the 1-tangle, and
(\ref{4q.1c}) the entanglement of $T$. (\ref{4q.1c}) is the
entanglement of formation if $T$ is a partial trace of a bipartite
quantum system. By its very definition we need only the values of
$\det T$ and of $S_T$ for pure input states.

There are some general relation between these three roofs. The first
is typical also for more general settings: $C_T$ is a positive
convex function and so does its square. For pure input states the
tangle and the squared concurrence coincide. Hence, because $\tau_T$
is maximal within all convex extensions, it is not less than
$C_T^2$. On the other hand, if $\omega$ turns out to be a flat point
of $C_T$, than this remains true for its \linebreak square. Thus,
\\[12pt]
\noindent {\bf Proposition 4.3}
\vspace{12pt}

For stochastic maps with 1-qubit outputs it holds
\begin{equation} \label{4q.2}
\tau_T(\omega) \geq C_T(\omega)^2
\end{equation}
and equality takes place for flat roof points of $C_T$.

Let us now switch to $S_T$. With use the abbreviations
$\eta(x) = - x \log x$ and
\begin{equation} \label{4q.3}
\xi(x) =
\eta \left(\frac{1-y}{2}\right) + \eta \left(\frac{1+y}{2}\right)
, \quad 1 = x^2 + y^2
\end{equation}
$\xi$ is defined and continuous  for $-1 \leq x \leq 1$ and
it is strictly convex. Therefore, $\xi$ is the sup of a
family of functions $ax + b$. Inserting a convex function
$C$ defined on any convex set with values
$-1 \leq C \leq 1$, we get $\xi(C)$ by a sup of
convex functions $aC + b$. Therefore, $\xi(C)$
is a convex function on the domain of definition of $C$.

We apply this fact to the concurrence $C_T$ yielding:
\begin{equation} \label{4q.4}
\omega \to \xi_T(\omega) := \xi(\, C_T(\omega) \,)
\end{equation}
is a convex function on $\Omega^d$. $C_T$ for pure states $\pi$
we get $2\sqrt{\det T(\pi)}$. We insert in (\ref{4q.4}),
\begin{displaymath}
\xi_T(\pi) = \eta(\frac{1-\sqrt{1 - 4 \det T(\pi)}}{2}) +
\eta(\frac{1+\sqrt{1 - 4 \det T(\pi)}}{2})
\end{displaymath}
and one identifies the arguments in $\eta$ as the two eigenvalues
of $T(\pi)$.
\begin{equation} \label{4q.5}
\xi_T(\pi) = S(T(\pi)), \quad \pi \in \Omega^{d, {\rm pure}} \;
\end{equation}
proves $\xi_T$ to be a convex extension of the pure states output
entropies. Reasoning as for proposition 4.3 results in
\\[12pt]
\noindent {\bf Proposition 4.4}
\vspace{12pt}

For stochastic maps with 1-qubit outputs it holds
\begin{equation} \label{4q.6}
E_T(\omega) \geq \xi(\, C_T(\omega) \,) \;
\end{equation}
Equality takes place if $\omega$ is a flat roof point.

It should be underlined hat there are more and different estimations
for concurrence and entanglement of formation in higher dimensions,
see \cite{CAF05a,CAF05b,LBZW03,MKB04,Os06}. In \cite{FJLW03} there
is an application to states with only two different, but arbitrarily
degenerated eigenvalues.

\subsection{How to Find $\vartheta$}
There is a general recipe to get the wanted anti-linear
operator for channels $T$ mapping the states of a
quantum system $\cH_d$ to 1-qubit states. Assume a Kraus
representation
\begin{equation} \label{kraus1}
T(X) = \sum A_j X A_j^{\dag}, \qquad
A_j \, : \quad \cH_d \mapsto \cH_2 \;
\end{equation}
There is an additional condition
to be fulfilled: $T$ must be Kraus representable with not more than
two Kraus operators. But at first we remain within the more
general (\ref{kraus1}).

The key to the following is the existence of the ``time reversal''
or ``spin flip'' anti-unitary operator $\theta_{\rm f}$ on $\cH_2$,
\begin{equation} \label{flip1}
\theta_{\rm f} (c_0 |0\rangle + c_1 |1\rangle) =
c_1^* |0\rangle - c_0^* |1\rangle \;
\end{equation}
Apart from the obvious
\begin{displaymath}
\theta_{\rm f}^{\dag} = \theta_{\rm f}^{-1} = - \theta_{\rm f}
\end{displaymath}
the relation
\begin{equation} \label{flip2}
\theta_{\rm f} Y^{\dag} \theta_{\rm f} Y = - (\det Y) \, \1_2
\end{equation}
is valid for all operators $Y \in \cB(\cH_2)$. Up to a
multiplicative constant, only the spin flip commutes with all $U
\in$ SU(2), It is really a very special anti-unitary operator.

The task is in inserting $Y = T(X)$ into (\ref{flip2}) and to
get something similar for $\det T$. This goes through
particulary nice if $X$ is of rank one,
$X = |\psi_2 \rangle\langle \psi_1|$.
Calculations show
\begin{equation} \label{flip3}
\det T(|\psi_2 \rangle\langle \psi_1 |) =
\sum_{j < k}  \, \langle \psi_1 , \vartheta_{jk} \psi_1 \rangle  \,
\langle \psi_2 , \vartheta_{jk} \psi_2 \rangle^* ,
\quad |\psi_i\rangle \in \cH_d
\end{equation}

The anti-linear Hermitian operators $\vartheta_{jk}$ are
defined by
\begin{equation} \label{flip4}
\vartheta_{jk} = {1 \over 2} \bigl(
A_j^* \theta_{\rm f} A_k - A_k^* \theta_{\rm f} A_j \bigr) \;
\end{equation}
using the Kraus operators $A_j$ from any Kraus representation
(\ref{kraus1}) of $T$. The operators $\vartheta_{jk}$ are
Hermitian and anti-linear.

In the lucky case of channels (\ref{kraus1}) with just two
Kraus operators, $A_1$ and $A_2$, we get only one operator
$\vartheta$ by (\ref{flip2}) and, hence,
\begin{equation} \label{flip5}
\sqrt{ \det T(|\psi \rangle \langle \psi |)} =
| \langle \psi, \vartheta \, \psi \rangle |
\end{equation}
and, by proposition 4.1, we are done.

Wether and how one can replace the operation
$X \to \vartheta X \vartheta$ by an anti-linear stochastic
map to obtain a more general roof construction, is unknown.
An (implicit) attempt
is in \cite{RBCHMW} by taking $X \to (\T \,X )\1 - X$ as an
higher dimensional substitute for the flip operation.

\subsection{Applications}
The partial trace of a 2-quibt system can be represented by two
Kraus operators: Looking at the operators over
$\cH_2 \otimes \cH_2$ as block matrices, the partial trace
over the second part is the map
\begin{equation} \label{wap1}
 X =  \pmatrix{X_{00} & X_{01} \cr X_{10} & X_{11}}
\, \longrightarrow \,    X_{00} + X_{11} = Y
\end{equation}

One immediately sees a possible choice for the Kraus operators,
\begin{equation} \label{wap2}
A_1 = \frac{1}{\sqrt{2}} \{ \1_2, \, \0_2 \}, \quad
A_2 = \frac{1}{\sqrt{2}} \{ \0_2, \, \1_2 \} \;
\end{equation}

Now we can compute $\vartheta$ according to (\ref{flip4}).
We get, eventually up to a sign, Wootters' conjugation,
$\vartheta = \theta_{\rm w}$,
\begin{equation} \label{wap3}
4 \vartheta = \theta_{\rm w} =
\pmatrix{ \0 & \theta_{\rm f} \cr - \theta_{\rm f} & \0 }
= \theta_{\rm f} \otimes \theta_{\rm f}
\end{equation}

Remark: The concurrence of a 2-qubit system is a flat
convex roof. Generally, complete convex leaves consists
of a set of flat ones.

Let us choose two operators,  $A_1 = A$, $A_2 = B$, from
$\cB(\cH_2)$ so that (\ref{kraus1}) becomes a 1-qubit channel.
Then there are only two eigenvalues, $\lambda_1$, $\lambda_2$,
of $|\sqrt{\omega} \vartheta \sqrt{\omega}|$ to respect and
(\ref{aroofth1a}) simplifies to
\begin{equation} \label{4q.7}
g^{\cup}(\omega) = |\lambda_2 - \lambda_1 |
\end{equation}

One has to solve a quadratic equation to get the general
expression
\begin{equation} \label{4q.8}
{1 \over 4} \, C_T(\omega)^2 =
\T \, (\omega \vartheta \omega \vartheta)
- 2 (\det X) \, (\det \vartheta^2)^{1/2}
\end{equation}

There are standard forms for 1-qubit channels,
\cite{GS76,FA98,RSW00,KR00}. For channels with two Kraus operators
one can assume
\begin{equation} \label{4q.9}
A = \pmatrix{a_{00} & 0 \cr 0 & a_{11}}, \quad
B = \pmatrix{0 & b_{01} \cr b_{10} & 0}
\end{equation}
up to unitary equivalence.
With this choice $\vartheta$ acts as
\begin{equation} \label{4q.10}
\vartheta (c_0 |0\rangle + c_1 |1\rangle) = (b_{10} a_{00} c_0)^*
|0\rangle - (b_{01} a_{11} c_1)^* |1\rangle \;
\end{equation}
After inserting in the relevant expressions one observes that
the concurrence is the restriction of a semi-norm to the
state space. Indeed, one obtains, \cite{Uh05,Uh03c},
\begin{equation} \label{4q.11}
C_T(X) = |\, |b_{10} a_{00}| x_{00} + |b_{01} a_{11}| x_{11}
+ z x_{10} - z^* x_{01} \,| \,
\end{equation}
$z$ is one of the roots of
\begin{equation} \label{4q.12}
 z^2 = a_{00}^* a_{11} b_{01} b_{10}^* \;
\end{equation}

\section{A Subtraction Procedure}
We start with a simple example not covered by Wootters' method
and showing the principle. The idea is to
subtract from $\det T(X)$ a suitable multiple of $\det(X)$
to get the squared concurrence or the tangle of a stochastic
1-qubit map $T$. We choose for $T$ the map
\begin{equation} \label{hh1}
\pmatrix{x_{00} & x_{01} \cr x_{10} & x_{11}}
 \, \longrightarrow \,
\pmatrix{x_{00} + (1-\gamma) x_{11} & 0 \cr 0 & \gamma x_{11} }
\end{equation}
and consider
\begin{equation} \label{hh2}
\gamma x_{00} x_{11} + \gamma (1- \gamma)x_{11}^2
- w (x_{00} x_{11} - x_{01} x_{10}) \;
\end{equation}
With $w = \gamma$, we arrive at the squared concurrence of $T$,
\begin{equation} \label{hh3}
\frac{1}{4} \,
C_T(X)^2 = \gamma (1-\gamma) x_{11}^2 + \gamma x_{01} x_{10}
\end{equation}
At first (\ref{hh3}) is a positive semi-definite quadratic form
implying that its square root is convex. Secondly, by its
very construction it coincides
for pure states $\pi$ with $4 \det T(\pi)$. Finally, on the state
space, it is a roof. To indicate a general way to proof the
roof property we polarize (\ref{hh3}) and get
\begin{equation} \label{hh4}
(X, Y)_T := \gamma (1-\gamma) x_{11} y_{11} + \frac{1}{2}
\gamma (x_{01} y_{10} + x_{10} y_{01})
\end{equation}
which is a positive semi-definite bilinear form in the space
of Hermitian matrices. For the pure state
$\pi_0 = |0 \rangle\langle 0|$ (\ref{hh3}) becomes zero.
Applying the Schwarz inequality we get $(X, \pi_0) = 0$ for
all $X$. Hence, with any pure state $\pi$,
\begin{equation} \label{hh5}
\omega_s = (1-s) \pi_0 + s \pi \, \Longrightarrow
(\omega_s, \omega_s) = s^2 (\pi, \pi) \,
\end{equation}
Taking the root we see that $s \to \omega_s$,
$0 \leq s \leq 1$, is a convex leaf of $C_T$ for every
$\pi$. Now the assertion is proved and, the more,
$C_T$ is {\em not} a flat roof.

One may ask, whether one can diminish $w$ a bit without destroying
convexity on $\Omega$. Let us use in (\ref{hh2}) the new value
$w' = \gamma^2$. Then (\ref{hh2}) becomes
\begin{displaymath}
\gamma x_{00} x_{11} + \gamma (1- \gamma)x_{11}^2
- \gamma^2 (x_{00} x_{11} - x_{01} x_{10}) \;
\end{displaymath}
Different to the former case we restrict ourselves to
$\Omega(\cH_2)$ and respect the condition
$x_{00} + x_{11} = 1$. After some
manipulations we arrive at a convex roof which coincides
with $\det T(\pi)$ for pure states. Up to a factor it must be
the tangle of $T$ on $\Omega$.
\begin{equation} \label{hh6}
\frac{1}{4} \tau_T(X) = \gamma (1-\gamma) x_{11} + \gamma^2
x_{01} x_{10} \;
\end{equation}
The tangle is affine on the set of density operators with $x_{01} =$
constant. As the square of the concurrence is equal to the
tangle for pure states, we have the inequality
$\tau_T(\omega) > C_T(\omega)^2$ for mixed states.

That the ansatz (\ref{hh2}) is working generally for the
concurrence has been shown first in
\cite{hildebrand07,hildebrand08,hellmund09}, by means of
the ``S-lemma of Yakobovich'' and, using the explicit expression
for general stochastic 1-qubit maps of \cite{GS76}, in \cite{HU08a}.
The case of the tangle can be read off from \cite{OV06}.

\subsection{Concurrence of Stochastic 1-Qubit Maps}
Let $T$ be a stochastic, \emph{i.e.}, a positive trace preserving
map. We prove
\\[12pt]
\noindent {\bf Proposition 5.1}
\vspace{12pt}

There is a real number $0 \leq w \leq 1$ such that for all
$\rho \in \Omega(\cH_2)$
\begin{equation} \label{hh7}
\frac{1}{4} \, C_T(\rho)^2 = \det T(\rho) - w \det \rho \;
\end{equation}

At first we show the $w$-bounds. With $w < 0$,
(\ref{hh7}) becomes the sum of two concave functions on $\Omega$
and $C_T$ could not be convex. To prove $1 \geq w$ we insert
$\rho_0 = (1/2) \1$ and get $\det T(\rho_0) \geq w/4$. However,
$\det T(\rho_0) \leq 1/4$ is required by stochasticity.
$w > 1$ would be a contradiction.

Next, we consider the expression (\ref{hh7}) on
the Bloch space of all Hermitian operators of trace one,
\begin{equation} \label{hh8}
X = \frac{1}{2}
(\1 + x_1 \sigma_1 + x_2 \sigma_2 + x_3 \sigma_3) \;
\end{equation}

The determinant of $X$ is a quadratic function
\begin{displaymath}
\det X = \frac{1}{4} ( 1 - x_1^2 - x_2^2 - x_3^2)
\end{displaymath}
in the Bloch coordinates. The same is with $\det T(X)$.
In the terminology of the S-lemma, the quadratic function
$\det T(X)$ is called {\em co-positive with} $\det X$ because
from $\det X \geq 0$ it follows $\det T(X) \geq 0$ under the
constraint $\T \, X = 1$.

The S-lemma, \cite{PT07}, states the following: Let $q_1$ and $q_2$
be two quadratic functions on $\bbbr^n$, not necessarily
homogeneous.  If $q_1(x) \geq 0$ implies $q_2(x) \geq 0$
then $q_2$ is co-positive with $q_1$.
\\[12pt]
\noindent {\bf S-lemma}
\vspace{12pt}

Let $q_1$ be strictly positive at least at one point. Then
there exists a real number $w$ such that
\begin{equation} \label{hh9}
q_2(x) - w q_1(x) \geq 0 \, \hbox{ for all } \, x \in \bbbr^n
\end{equation}
if and only if $q_2$ is co-positive with $q_1$.

Now it is proved: The right expression of (\ref{hh7}) can
be made non-negative with suitable
numbers $w$ for all Hermitian trace one operators.
Then, substituting $x_j \to x_j/x_0$ and multiplying
by $x_0^2$, we see that
\begin{equation} \label{hh10}
\det \T(X) - w \det X \geq 0
\end{equation}
becomes a positive semi-definite and homogeneous polynomial
in the variables $x_0, x_1, x_2, x_3$ for some $w$.
Polarizing (\ref{hh10}) as in (\ref{hh4}) yields a
positive semi-definite symmetric form $(X,Y)_w$ satisfying
\begin{equation} \label{hh11}
(X,X)_w = \det T(X) - w \det X \;
\end{equation}
We now may assume (\ref{hh10}) for all values $w$ which are
bounded by $1 \geq w_1 \geq w \geq w_2 \geq 0$.
We can further assume that $(X,Y)_w$ is degenerated for
$w = w_1$ and for $w = w_2$.
But wether degenerate or not, (\ref{hh10})
implies Cauchy's inequality
\begin{equation} \label{hh12}
|(X,Y)_w|^2 \leq (X,X)_w (Y,Y)_w \;
\end{equation}
In particular, if $Y = \nu$ is in the null-space,
$(\nu, \nu)_w = 0$, then $\det T(\nu) = w \det \nu$
and $\det T(\nu) \geq w' \det \nu$ for all
allowed $w'$, \emph{i.e.},
\begin{equation} \label{hh12a}
(w - w') \det \nu \geq 0, \quad w_1 \geq w' \geq w_2 \;
\end{equation}
Furthermore, $(\nu, X)_w = 0$ for all Hermitian $X$.

 We like to show: If $w = w_2$ every density operator
$\rho$ is a roof point.

We have to distinguish several cases. The first one is
$\T \, \nu = 0$. Then, with $\rho \in \Omega$, we define
\begin{equation} \label{case1}
\rho_s = \rho + s \nu \,
\end{equation}
On this line $(\rho_s, \rho_s)_W$ is independent of
$s$. Therefore, because of (\ref{hh7}), its intersection
with $\Omega$ is a flat convex leaf of $C_T$.

Being Hermitian and not identical zero, $\T \, \nu = 0$
results in $\det \nu < 0$. Because of (\ref{hh12a})
one concludes $w = w_2$.

For the next cases we suppose $\T \, \nu =1$ and
start with
\begin{equation} \label{case2}
\rho_s =  (1-s) \nu + s \rho  \,
\end{equation}
Then $(\rho_s, \rho_s)_W = s^2$.
If the sign of $s$ does not change while $\rho_s$
is inside the Bloch ball, we can take the root
and get a convex leaf. This takes place if $\nu$
is {\em not} an inner point of $\Omega$, \emph{i.e.}, not a
properly mixed state.

The condition is certainly satisfied if
$\det \nu = 0$ and $\nu$ is a pure state.
The condition implies the $w$-independence
of $(\nu, \nu)_w$ which becomes equal to
$\det T(\nu)$.  As there is only one convex roof
$C_T$ it is $w_1 = w_2$ necessarily \cite{fnote6}.

$\nu$ is outside $\Omega$ if $\det \nu \neq 0$.
We conclude $w = w_2$ from (\ref{hh12a}) and
we get convex leaves by intersecting the lines
(\ref{case2}) with the Bloch ball.

In addition one observes: $\det \nu \neq 0$ is necessary for $w_1 >
w_2$. Indeed, by (\ref{hh12a}) we see $(\nu, \nu)_w = 0$ can be
satisfied if either $w = w_1$ and $\det \nu > 0$ or by $w = w_2$ and
$\det \nu < 0$.

Now, all relevant cases are discussed and proposition 5.1
is proved.

In the course of the proof one obtains two more general
insights:
\\[12pt]
\noindent {\bf Proposition 5.2}
\vspace{12pt}

The concurrence of stochastic 1-qubit maps is the restriction
of a Hilbert semi-norm to the state space.\\
Every state allows for an optimal decomposition of length
two.

\subsection{Axial Symmetric Maps, Concurrence}
We shall list the subtraction parameter $w$ for the class
of axial symmetric stochastic maps. A standard form
for them reads
\begin{equation}   \label{axi.1}
  T(X) =
\begin{pmatrix}{
\alpha x_{00}  +(1-\gamma) x_{11}     & \beta x_{01} \cr
\beta x_{10} &   \gamma x_{11} + (1 - \alpha) x_{00} }
\end{pmatrix}
\end{equation}
with real non-negative parameters $\alpha,\beta,\gamma$.
The trace preserving is obvious. Positivity requires
$0 \leq \alpha \leq 1$, $0 \leq \gamma \leq 1$, and
\begin{equation} \label{axi.2}
      \beta^2 \leq \; \beta^2_{\max} := \;
1+2\alpha\gamma-\alpha-\gamma +
2\sqrt{\alpha(1-\alpha)\gamma(1-\gamma)} \;
\end{equation}
$T$ is a channel, hence completely positive, if
$\beta^2 \leq \alpha \gamma$.
To express $w$ one needs the ``critical'' $\beta_c$
\begin{equation} \label{axi.3}
\beta_c \; := \;
1+2\alpha\gamma-\alpha-\gamma -
2\sqrt{\alpha(1-\alpha)\gamma(1-\gamma)} \;
\end{equation}
Then, see \cite{hellmund09},
\begin{equation} \label{axi4}
 w = \max \{\beta^2, \beta_c^2\}
\end{equation}

At the bifurcation point $\beta = \beta_c$ the $T$-concurrence
is  affine on $\Omega$. For $\beta \geq \beta_c$ the roof is
flat. Otherwise it looks similar to the particular one
(\ref{hh1}). See \cite{hellmund09} for more details.

\subsection{Axial Symmetric Maps, Tangle}
The tangle for stochastic 1-qubit maps can be found in
\cite{hildebrand07,hildebrand08}. For 1-qubit channels
it is already \linebreak in \cite{OV06}. These tangles always
allow for optimal decompositions of length two.

The axial symmetric maps (\ref{axi.1}) can be treated explicitly.
The following is due to \cite{Hellm}.

\noindent Case A: \, If $|\beta| > |\alpha + \gamma -1|$ one
has to use $w = \beta^2$. It results in
\begin{equation} \label{axi5}
\frac{1}{4} \tau_T(X) =
1 - \beta^2 - (\alpha - \gamma)^2 -
2(\alpha - \gamma)(\alpha + \gamma -1)x_3
+ [\beta^2 - (\alpha - \gamma -1)^2]x_3^2
\end{equation}

\noindent Case B: \, Here $|\beta| = |\alpha + \gamma -1|$,
a bifurcation point in the parameter space. $w = \beta^2$
 results in
\begin{equation} \label{axi6}
\frac{1}{4} \tau_T(X) =
1 - \beta^2 - (\alpha - \gamma)^2 -
2(\alpha - \gamma)(\alpha + \gamma -1)x_3
\end{equation}
and the tangle becomes affine on the Bloch ball.

\noindent Case C: \, If $|\beta| < |\alpha + \gamma -1|$
then $w = (\alpha + \gamma -1)^2$ and we obtain
\begin{equation} \label{axi7}
\frac{1}{4} \tau_T(X) =
1 - (\alpha + \gamma -1)^2 - (\alpha - \beta)^2 -
2(\alpha - \gamma)(\alpha + \gamma -1)x_3 +
[(\alpha + \gamma -1)^2 - \beta^2](x_1^2 + x_2^2)
\end{equation}

\section{Symmetries}
The use of symmetries is almost obligatory in the treatment of
roofs. We present only a small, hopefully helpful, part of it,
mainly abstracted from \cite{BNU96,BNU99,BNU02,VW96,TV00}. See also
\cite{FL06}.

Let $\Omega(\cH)$ be the space of states supported by the
Hilbert space $\cH$ of dimension $d$, and $g$ a  real continuous
function on $\Omega^{\rm pure}$.

A symmetry of $\Omega$ is a transformation
\begin{equation} \label{6.1}
\omega \to \omega^V := V \omega V^{-1} \;
\end{equation}
$V$ is a unitary or an anti-unitary operator {\em inducing}
the symmetry (\ref{6.1}).

We need the group $\Gamma$ of all $V$ such that
\begin{equation} \label{6.2}
\pi \in \Omega^{\rm pure} \, \Longrightarrow \,
g(\pi^V) = g(\pi) \;
\end{equation}
$\Gamma$ is the {\em invariance group of $g$.} A quite obvious
statement reads
\\[12pt]
\noindent {\bf Proposition 6.1:}
\vspace{12pt}

If $\Gamma$ is the invariance group of $g$, then
\begin{equation} \label{6.3}
g^{\cup}(\omega^V) = g^{\cup}(\omega), \quad
g^{\cap}(\omega^V) = g^{\cap}(\omega)
\end{equation}
for all $V \in \Gamma$ and all $\omega \in \Omega$.

Let $K$ be a convex leaf of $g^{\cup}$. Obviously,
\begin{equation} \label{6.4}
K^{V} = \{\omega^V \, | \, \omega \in K\}
\end{equation}
is a convex leaf of $g^{\cup}$ again. Hence, $K \to K^V$
permutes the convex leaves.

An interesting subgroup of $\Gamma$ is the stabilizer group
\begin{equation} \label{6.5}
\Gamma_K = \{ V \in \Gamma \, | \, K^V = K \} \;
\end{equation}
This is a compact group with an invariant Haar measure.
say $d_K V$.
We can perform the invariant integration (``twirling'')
over $\Gamma_K$,
\begin{equation} \label{6.6}
\omega \longrightarrow \omega^K = \int \omega^V d_K V \;
\end{equation}
There is only one $\Gamma_K$-invariant element in the
convex hull of all $\omega^V$. It is $\omega^K$.

The map $\omega \to \omega^K$ contracts $K$ onto the set
\begin{equation} \label{6.7}
K^{\rm stable} =
\{ \omega \in K \, | \, \omega = \int \omega^V d_K V \} \;
\end{equation}
This set, being convex and compact, is the convex hull of its {\em
extremal invariant states.} Extremal invariant states can be
represented by $\pi^K$ with pure $\pi \in K$. Invariant states which
are not extremal, cannot be represented in such a way.
\\[12pt]
\noindent {\bf Proposition 6.2:}
\vspace{12pt}

Let $K$ be a convex leaf of $g^{\cup}$ and $\Gamma_K$ its
stabilizer group. Then $K \cap \Omega^{\rm pure}$
consists of $\Gamma_K$-orbits.

Every extremal $\Gamma_K$-invariant state of $K$
is of the form $\pi^K$, $\pi$ pure.

Every $\Gamma_K$-invariant states of $K$ is a convex combination
of extremal $\Gamma_K$-invariant states.

\subsection{Entanglement of the Diagonal Channel}
In example 3.2 we considered the concave roof of (\ref{3e.3}),
\begin{displaymath}
g(\pi) = S({\rm diag}(\pi)), \quad \pi \in \Omega(\cH)
\end{displaymath}
Now we look at the convex roof $g^{\cup}$, \emph{i.e.}, at the
entanglement $E_D$ of $D(\omega) = {\rm diag}(\omega)$.

With the exception of $d=2$ one does not know the structure
of $E_D$. But there are some insights on highly symmetric,
``isotropic'' quantum states.

The channel $D$ will be described by the help of a basis
$|j\rangle$, $j=1, \dots, d$. To it we associate the vector
\begin{equation} \label{6.8}
|\psi \rangle = \frac{1}{\sqrt{d}} \sum |j\rangle \;
\end{equation}
The invariance group, $\Gamma$, of (\ref{6.8}) consists of the
permutations of the chosen basis, eventually followed by the
conjugation $\Theta$ defined by $\Theta |j\rangle = |j\rangle$ for
all $j$. The density operator $\omega$ commutes with $\Gamma$ if its
matrix elements satisfy
\begin{equation} \label{6.9}
\langle j| \omega| j \rangle = \frac{1}{d},
\quad
\langle j|\omega|k \rangle = \frac{x}{d}
\end{equation}
for all $j$ and all $k \neq j$. $x$ is a real number in the
range
\begin{equation} \label{6.10}
- \frac{1}{d-1} \leq x \leq 1 \;
\end{equation}
The restriction is due to the positivity of $\omega$. One often
uses the {\em fidelity parameter,} $F$,
\begin{equation} \label{6.11}
0 \leq F := \langle \psi| \omega |\psi\rangle =
\frac{(d-1)x +1}{d-1} \leq 1
\end{equation}
and we denote the corresponding $\Gamma$-invariant density
operator by $\omega_F$.

We choose an allowed value $F$ and write $K_{\omega}$
or $K(F)$ for the complete convex leaf of $\omega_F$ with
respect to $E_D$. Now we state the following:
\begin{equation} \label{6.12}
\rho \in K_{\omega} \, \Rightarrow \,
\Theta \rho = \rho \Theta \;
\end{equation}
$\Theta$ is the conjugation about the basis $\{|j\rangle\}$.\\
Proof:
It suffices to prove the assertion for pure states. We
assume that the pure states $\pi$ and $\pi' = \Theta \pi \Theta$
are both in $K_{\omega}$. The diagonal parts of them and of $\rho
= (1/2)(\pi + \pi')$ are the same. It follows $S_D(\rho) =
S_D(\pi) = S_D(\pi')$ and, because we are inside a convex leaf, we
get also $g^{\cup}(\rho) = S_D(\rho)$. Hence, by proposition 3.2,
$S_D = E_D$ on the whole face containing $\pi$ and $\pi'$. This
is a contradiction if $\pi \neq \pi'$.

Indeed, we proved something more:
\\[12pt]
\noindent {\bf Proposition 6.3:}
\vspace{12pt}

Any two different pure states contained in a convex leaf
of $E_D$ must have different diagonal parts.

The $\Gamma$-invariant density operators are ordered by
the fidelity parameters as indicated by (\ref{6.11}).
The more, $F \to \omega_F$ is convexly linear in $F$.
This fact allows to apply proposition 6.2.
\\[12pt]
\noindent {\bf Proposition 6.4:}
\vspace{12pt}

Let $K(F)$ denote the compact convex leaf of $E_D$ belonging
to the $\Gamma$-invariant density operator $\omega_F$.
There is a subset $R$ of the unit interval as follows:\\
a) Either we have $F \in R$. Then $K(F)$ consists of flat roof
points only. There is no other maximal
symmetrical state in $K(F)$ than $\omega_F$.\\
b) Or there are $F^{-}, F^{+} \in R$ such that $K(F)$ is the
convex hull of $K(F^-) \cap K(F^+)$.

In the case $d=3$ and if
\begin{displaymath}
\frac{1}{2} \leq F \leq F^{**}, \quad F^{**} = \frac8{9}
\end{displaymath}
or $F = 0$ case a) is true. There is an optimal vector of the form
$a |1\rangle + b |2\rangle + b |3\rangle$ for any of these
$F$-values. $K_{\omega} = K(F)$ contains not more than three pure
states. They become permuted by the action of the invariance group
$\Gamma$.

For more details, also in higher dimensions, see \cite{BNU02}.

\subsection{An Embedding}
There are numerous relations between different channels. Some of
them provide insight in roof structures. Our main interest is
again in the diagonal channels.

Let $\cH$ and $\cH'$ be Hilbert spaces of dimensions $d$ and $d' >
d$. There are embeddings of $\cH$ into $\cH'$ which relate the
entanglement $E_D$ and $E_{D'}$ of the corresponding diagonal maps,
$D$ and $D'$.

For their description we first choose $d$ integers, $m_1, \dots,
m_d$, such that $d' = m_1 + \dots + m_d$ and enumerate the basis
vectors of $\cH'$ as
\begin{equation} \label{6e.1}
|jk\rangle, \quad j = 1, \dots, d ,
\quad k = 1, \dots, m_j \;
\end{equation}
We further choose numbers
\begin{equation} \label{6e.2}
y_{j,k}, \quad j = 1, \dots, d ,
\quad k = 1, \dots, m_j
\end{equation}
satisfying
\begin{equation} \label{6e.3}
\sum_{k=1}^{m_j} |y_{jk}|^2 = 1 , \quad j = 1, \dots, d \;
\end{equation}
These data provide a unitary embedding
\begin{equation} \label{6e.4}
|j\rangle \to V |j\rangle =
\sum_{j=1}^d y_{jk} |j,k\rangle \;
\end{equation}
From
\begin{equation} \label{6e.5}
{\rm diag}(X) = \{x_{11}, \dots, x_{dd}\}
\end{equation}
we get
\begin{equation} \label{6e.6}
{\rm diag}(VXV^{\dag}) = \{|y_{11}|^2 x_{11}, \dots,
|y_{1m_{1}}|^2 x_{11}, |y_{21}|^2 x_{22},
\dots |y_{2m_{1}}|^2 x_{22}, \dots \}
\end{equation}
One obtains for the entropy of the diagonal channel
\begin{equation} \label{6e.7}
S_{D'}(VXV^{\dag}) =
\sum_{k=1}^{m_1} \eta(|y_{1k}|^2 x_{11}) +
\sum_{k=1}^{m_2} \eta(|y_{2k}|^2 x_{22}) + \dots
\end{equation}
The functional equation $\eta(xy) = y \eta(x) + x \eta(y)$
allows to rewrite the first sum in (\ref{6e.7}) into the form
\begin{displaymath}
\sum |y_{1k}|^2 \eta(x_{11}) + x_{11}
\sum \eta(|y_{1k}|^2) \;
\end{displaymath}
and we get finally
\begin{equation} \label{6e.8}
S_{D'}(V \omega V^{\dag}) = S_D(\omega) +
\sum_{j=1}^m \langle j| \omega |j\rangle
\sum_{k=1}^{m_j} \eta(|y_{jk}|^2) \;
\end{equation}
A convex roof remains a convex roof if we add a function
linear in $\omega$. Plugging (\ref{6e.8}) into (\ref{n3.4a})
\linebreak directly provides:
\\[12pt]
\noindent {\bf Proposition 6.5:}
\vspace{12pt}

If $\cH$ is embedded unitarily in $\cH'$ according to (\ref{6e.4}),
then the entanglements of the diagonal channels are related by
\begin{equation} \label{6e.9}
E_{D'}(V \omega V^{\dag}) = E_D(\omega) + l(\omega)
\end{equation}
and the linear function is given by
\begin{equation} \label{6e.10}
l(\omega) =
\sum_{j=1}^m \langle j| \omega |j\rangle
\sum_{k=1}^{m_j} \eta(|y_{jk}|^2) \;
\end{equation}
Optimal decompositions are mapped onto optimal decomposition
and convex leaves onto convex leaves.

A particular simple example is the embedding
\begin{equation} \label{6e.11}
V |0\rangle = |1\rangle, \quad V |1\rangle =
\frac{1}{\sqrt{2}} (|2\rangle + |3\rangle)
\end{equation}
of $\cH_2$ into $\cH_3$. Pairs of pure state vectors,
yielding flat optimal decompositions for the
entanglement of $D_2$, are
\begin{equation} \label{6e.12}
a_0 |0\rangle + a_1 |1\rangle, \quad
a_1^{*} |0\rangle + a_0^* |1\rangle \;
\end{equation}
A particular case is $a_0 =ÿ\sqrt{1/3}$, $a_1 = \sqrt{2/3}$.
Applying the map (\ref{6e.11}) results in the optimal pair
\begin{equation} \label{6e.13}
\sqrt{\frac{1}{3}} (|1\rangle + |2\rangle + |3\rangle)
 , \quad \sqrt{\frac{2}{3}} |1\rangle +
\sqrt{\frac{1}{6}} (|2\rangle + |3\rangle)
\end{equation}
Call $\pi_0$ and $\pi_1$ the pure states determined
by the vectors (\ref{6e.13}). Then
\begin{equation} \label{6e.14}
{\rm diag}(\pi_0) = \{\frac{1}{3}, \frac{1}{3}, \frac{1}{3} \}
 , \quad
{\rm diag}(\pi_1) = \{\frac{2}{3}, \frac{1}{6}, \frac{1}{6} \}
\end{equation}
and the fidelity parameters (\ref{6.11}) are 1 and $F^{**} = 8/9$.
Returning to the remarks after proposition 6.4, we see a reason why
this value should be a bifurcation point for the behavior of maximal
symmetric states and their convex leaves.

\subsection{A Further Embedding}
There are strong relations between the entanglement of the
diagonal channels and the entanglement of formation, governed
by embedding procedures.
A nice and quite simple one is
\begin{equation} \label{6f1}
V |j\rangle = |jj\rangle = |j\rangle \otimes |j\rangle \;
\end{equation}
$V$ is a unitary map of $\cH$ onto the subspace $\cH'$
of $\cH^{a} \otimes \cH^{b}$ with basis $\{|jj\rangle\}$.
The embedding is of interest because of
\begin{equation} \label{6f2}
D(X) := {\rm diag}(X) = \T_b  (V X V^{\dag}),
\quad X \in \cB(\cH) \;
\end{equation}
The relation implies
\begin{equation} \label{6f3}
E((V \omega V^{\dag})) = E_D(\omega),
\quad \omega \in \Omega(\cH)
\end{equation}
and relates the entanglement $E_D$ of a diagonal map to the
entanglement of formation $E$ for bipartite states supported
by $\cH'$. This is true for all dimensions $d = \dim \cH$.
The vector (\ref{6.8}) becomes a completely entangled
one, say $|e\rangle$, if transformed by (\ref{6f1}).

The crucial point is now that invariance group $\Gamma_e$
of $|e\rangle = V |\psi \rangle$ is much larger than
$V \Gamma V^{\dag}$, the invariance group of
$|\psi\rangle$. $\Gamma_e$ contains the local unitary operators
$U \otimes \bar U$ and the swap operation. The involution
$V \Theta V^{\dag} = \Theta \otimes \Theta$ is
defined originally only on $\cH'$.
The canonical extension to an involution
$\Theta_e$ of all $\cH^{a} \otimes \cH^b$ can be  gained by
\begin{equation} \label{6f.4}
\Theta_e |jk\rangle = |kj\rangle
\end{equation}
and the requirement of anti-linearity. $\Theta_e$ satisfies
\begin{equation} \label{6f.5}
\Theta_e (X \otimes \1) |e\rangle = (X^{\dag} \otimes \1) |e\rangle
\end{equation}
for all $X \in \cB(\cH^{a})$ \cite{fnote5}.

Now one tries to enlarge convex leaves in $\cH'$ by transforming
them with operators from a suitable larger group $\Gamma' \subset
\Gamma_e$. The involution $\Theta_e$ is a symmetry of the
entanglement of formation. If a bipartite state $\rho$ commutes with
$\Theta_e$, all elements of the convex leaf of $\rho$ must commute
with $\Theta_e$.

One can understand quite well why $E_D$ for maximal symmetric states
are so similar to the entanglement of formation for isotropic
states. A detailed discussion is not in the frame of the present
paper. However, an essential point in the considerations above is in
the relation between the entanglement of diagonal channels and the
entanglement of formation. This may be of use in future research: It
seems easier to imagine the structure of the diagonal channel as
that of the partial trace. Nevertheless, the degree of difficulty is
about the same.

\section{Summary and Outlook}
Given values $g(\pi)$ for pure states $\pi$, the direct way of
solving the convex roof problem is the search for optimal
decompositions. The most prominent and successful examples are the
concurrence and the entanglement of formation for 2-qubit bipartite
quantum systems. The method goes back to Wootters and is described
in section 4. With it one gets analytical expressions and flat
optimal decompositions. The flatness of the convex (and concave)
roofs inherited from Wootters' method is rendering its use in higher
dimensions.

A quite different way is to look for a maximal convex extension $G$
of $g$ : If for any other extension $G'$ of $g$ we find $G'(\omega)
> G(\omega)$ for a state $\omega \in \Omega$, then $G'$ cannot be
convex.

A further, and more efficient reformulation of the convex roof
problem asks for roof points, (see definition 2.1a), of a
convex extension $G$ of $g$. At a roof point $\omega$ of
a convex extension $G$ one gets $G(\omega) = g^{\cup}(\omega)$
and the problem is solved for the particular state $\omega$.
Similarly, if $G$ would be a concave extension of $g$, then
$G(\omega) = g^{\cap}(\omega)$ for a roof point of $G$.

This way of proving is used in the chapter on the ``subtraction
procedure''. One of its merits is the control on the concurrence and
on the 1-tangle of any rank two density operator of a $2 \times m$
bipartite quantum system. The same is with the slightly more general
class of stochastic, (just positive and trace preserving), 1-qubit
channels.

Therefore, there is some hope to get the concurrence (and the
1-tangle) for all states of any $2 \times m$ bipartite quantum
system explicitly.

But even if this becomes true, it does not provide us with the
entanglement of formation of a $2 \times 3$ system: The concurrence
ceases to be flat. However, by proposition 4.1 one can obtain
reasonable \linebreak lower bounds.

Wootters' and the subtraction method seem to be quite different in
spirit. Uniting the strength of both, would
be very useful. Also one should look at the subtraction method in
more general terms. One can get at least lower bounds on the
concurrence for higher dimensional system as shown by two examples
in \cite{hellmund09}. A more systematic study of the issue seems
prospective. 

The use of symmetries is well know and efficient in general.
In convex roof construction the symmetries of $g$ as of a
function on the pure states is what counts. If a state
 $\omega$ is invariant with respect to a symmetry group
$\Gamma$, its convex leaf, (see definitions 3.2), is the convex hull
of a set of $\Gamma$-orbits consisting of pure states. The shapes of
the leaves can be quite different. However, one would suppose a
smooth change of the leaves with the exception of some bifurcation
points (or lines ...) at which the dimension of the leaves is
changing. Some help comes from embedding a lower dimensional problem
into a higher dimensional one. This is shown for the entanglement of
the diagonal channel: $E_T$ can be computed in any dimension on
2-dimensional subspaces which contain at least one pure diagonal
state. On these subspaces $E_T$ is the sum of a flat convex roof and
a linear  function. The study of more examples is certainly
desirable.

At this point we can return to the concurrence of the stochastic
1-qubit maps. For them $C_T$ is the restriction of a Hilbertian
semi-norm to the state space. The proof of proposition 5.1 shows a
one to one correspondence between the structure of the foliation and
the null-space of the semi-norm. Two stochastic 1-qubit maps come
with the same pattern of their convex leaves if the null-spaces of
their semi-norms are identical. Indeed, it is the first class of
channels and roofs with a complete classification of their convex
leaves and their foliation.

\section*{Acknowledgements}
I like to thank M.~Hellmund for many valuable discussions.
I wish to acknowledge the helpful and valuable advices of the
referees and of P.~Harremo\"es, Editor-in Chief of Entropy.


\bibliographystyle{mdpi}
\makeatletter
\renewcommand\@biblabel[1]{#1. }
\makeatother

\begin{thebibliography}{999}
\bibitem{fnote6}
This does not affect the 1-tangle.
Though (93) results from a subtraction procedure,
it is not an application of the S-lemma:
$\tau_T$ becomes negatve for $x_{00} \to \infty$.
\bibitem{fnote7}
If the null space is 2-dimensional the convex leaves
of $C_T$ become 2-dimensional too.
A 3-dimensional null-space
forces $C_T$ to be constant.
\bibitem{fnote8}
A flat roof point can allow for several optimal
decompositions, some of them flat and some of them
not flat.
\bibitem{fnote1}
This is the
very reason the present author introduced the notation
``convex roof''.
\bibitem{fnote2}
If $g$ is not continuous, an optimal decomposition may not exist,
even if we allow for infinite length.
\bibitem{fnote3}
Anti-linear Operators are also called a ``conjugate linear''
ones.
\bibitem{fnote4}
The polar decomposition for anti-linear operators
is equivalent to the Takaki decomposing of
symmetric matrices.
\bibitem{fnote5}
$\Theta_e$ is the so-called modular conjugation with defining
vector $|e\rangle$.
\end{thebibliography}

\end{document}